\documentclass[aps,prx,superscriptaddress,amsmath,amssymb,twocolumn,showpacs,floatfix,english,noshowpacs,longbibliography]{revtex4-1}

\usepackage{url}
\usepackage{graphicx}
\usepackage{color}
\usepackage[colorlinks=true, urlcolor=blue, linkcolor=blue, citecolor=blue, pdftex]{hyperref}
\usepackage[english]{babel}
\usepackage{braket}
\usepackage{amssymb}
\usepackage{subfigure}

\DeclareGraphicsExtensions{.pdf,.png,.jpg}

\newcommand{\ham}{\hat{\mathcal{H}}}

\newcommand{\sample}{Ba(TiO)Cu$_4$(PO$_4$)$_4$}

\usepackage{graphicx}
\usepackage{dcolumn}
\usepackage{bm}

\begin{document}

\title{Spin dynamics in the square-lattice cupola system 
Ba(TiO)Cu$_4$(PO$_4$)$_4$}

\author{Luc Testa}
\email[]{luc.testa@gmail.com}
\affiliation{Institute of Physics, Ecole 
Polytechnique F\'ed\'erale de Lausanne (EPFL), CH-1015 Lausanne, Switzerland}

\author{Peter Babkevich}
\affiliation{Institute of Physics, Ecole 
Polytechnique F\'ed\'erale de Lausanne (EPFL), CH-1015 Lausanne, Switzerland}

\author{Yasuyuki Kato}
\affiliation{Department of Applied Physics, University of Tokyo, Hongo, 
Tokyo 113-8656, Japan}

\author{Kenta Kimura}
\affiliation{Department of Advanced Materials Science, University of Tokyo, 
Kashiwa, Chiba 277-8561, Japan}

\author{Virgile Favre}
\affiliation{Institute of Physics, Ecole 
Polytechnique F\'ed\'erale de Lausanne (EPFL), CH-1015 Lausanne, Switzerland}

\author{Jose A. Rodriguez-Rivera}
\affiliation{NIST Center for Neutron Research, National Institute of 
Standards and Technology, Gaithersburg, MD 20899, USA}
\affiliation{Materials Science and Engineering Department, University of 
Maryland, College Park, MD 20742, USA}

\author{Jacques Ollivier}
\affiliation{Institut Laue-Langevin, BP 156, 38042 Grenoble Cedex 9, France}

\author{St\'ephane Raymond}
\affiliation{Universit\'e Grenoble Alpes, CEA, IRIG, MEM, MDN, 38000 
Grenoble, France}

\author{Tsuyoshi Kimura}
\affiliation{Department of Advanced Materials Science, University of Tokyo, 
Kashiwa, Chiba 277-8561, Japan}

\author{Yukitoshi Motome}
\affiliation{Department of Applied Physics, University of Tokyo, Hongo, 
Tokyo 113-8656, Japan}

\author{Bruce Normand}
\affiliation{Institute of Physics, Ecole 
Polytechnique F\'ed\'erale de Lausanne (EPFL), CH-1015 Lausanne, Switzerland}
\affiliation{Paul Scherrer Institute, CH-5232 Villigen PSI, Switzerland}

\author{Henrik M. R\o nnow  }
\email[]{henrik.ronnow@epfl.ch}
\affiliation{Institute of Physics, Ecole 
Polytechnique F\'ed\'erale de Lausanne (EPFL), CH-1015 Lausanne, Switzerland}

\date{\today}

\begin{abstract}
We report high-resolution single-crystal inelastic neutron scattering 
measurements on the spin-1/2 antiferromagnet Ba(TiO)Cu$_4$(PO$_4$)$_4$. 
This material is formed from layers of four-site ``cupola'' structures, 
oriented alternately upwards and downwards, which constitute a rather 
special realization of two-dimensional (2D) square-lattice magnetism. 
The strong Dzyaloshinskii-Moriya (DM) interaction within each cupola, 
or plaquette, unit has a geometry largely unexplored among the numerous 
studies of magnetic properties in 2D Heisenberg models with spin and spatial 
anisotropies. We have measured the magnetic excitations at zero field and 
in fields up to 5~T, finding a complex mode structure with multiple 
characteristic features that allow us to extract all the relevant magnetic 
interactions by modelling within the linear spin-wave approximation. We 
demonstrate that Ba(TiO)Cu$_4$(PO$_4$)$_4$ is a checkerboard system with almost 
equal intra- and inter-plaquette couplings, in which the intra-plaquette DM 
interaction is instrumental both in enforcing robust magnetic order and in 
opening a large gap at the Brillouin-zone center. We place our observations 
in the perspective of generalized phase diagrams for spin-1/2 square-lattice 
models and materials, where exploring anisotropies and frustration 
as routes to quantum disorder remains a frontier research problem. 
\end{abstract}

\maketitle

\section{Introduction}
\label{si}

Despite its apparent simplicity, the antiferromagnetic (AF) spin-1/2 Heisenberg 
model encapsulates all the rich many-body physics of non-commuting quantum 
variables. Even on a one-dimensional (1D) chain with only nearest-neighbor 
interactions, its exact solution describes a strongly fluctuating spin state 
with fractionalized excitations \cite{Bethe_1931,Faddeev_1981,Muller_1981}. 
On the square lattice in 2D, the nearest-neighbor ($J_1$) model shows 
spontaneous breaking of the continuous spin symmetry and N\'eel-type magnetic 
order \cite{Neel_1948,Shull_1951}, albeit with a quantum renormalization of 
the ordered moment to 61\% of its maximal value \cite{Manousakis_1991}. The 
idea that quantum fluctuations could destroy this order in 2D was put forward 
originally for the triangular lattice \cite{Anderson_1973}, on which (AF) 
interactions are geometrically frustrated. While the concept of the resonating 
valence bond (RVB) state was not realized on this lattice, it returned to 
prominence in the context of cuprate superconductivity \cite{Lee2006}, and is 
a leading candidate for the ground state of the square lattice frustrated by 
diagonal next-nearest-neighbor ($J_2$) interactions \cite{Chandra_1988}. In 
this sense, frustrated 2D Heisenberg models are the original prototype for 
quantum spin-liquid states \cite{Balents_2010}. 

\begin{figure*}[t]
\centering
\includegraphics[width=0.96\textwidth]{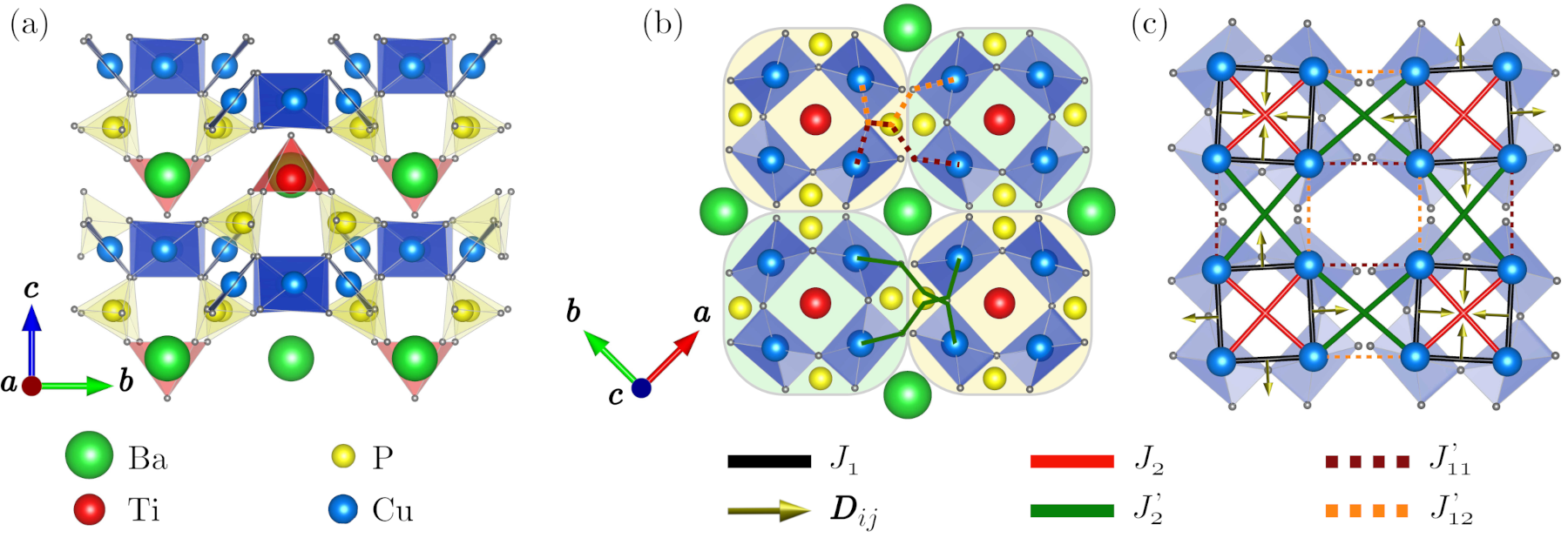}
\caption{\label{fig:structure} Schematic representation of the structure of 
Ba(TiO)Cu$_4$(PO$_4$)$_4$, showing Cu (blue), Ba (green), Ti (red), P (yellow), 
and O (grey) atoms with associated coordination polyhedra. (a) Projection on 
the $ac$ plane, providing a side view of the buckled layers. (b) Projection 
on the $ab$ plane of the layers, highlighting the square cupola structures 
as four CuO$_4$ squares (blue) connected around a Ti atom. Yellow and green 
shading indicate respectively upward- and downward-oriented cupolas. (c) 
Representation of the square-planar ($ab$) magnetic lattice, meaning the 
interactions between Cu atoms defined in Eq.~\eqref{eq:hamiltonian}. Figure 
\ref{fig_magDM} provides perspective views of the cupola structures and the 
resulting geometry of the DM vectors, $D_{ij}$.}
\end{figure*}

The study of more complex square-lattice quantum antiferromagnets has been 
pursued in a number of directions in recent years. In the direction of 
frustration, both analytical and numerical studies of the $J_1$-$J_2$ model 
\cite{Isaev_2009_MF_square,Gotze_2012,Doretto2014,Gong2014,Morita2015,Wang2018,
Haghshenas2018} have reached a very high level of sophistication without 
reaching a consensus on the nature of the quantum disordered phases around 
$J_2 = J_1/2$. In the direction of spatial anisotropy, numerical investigation 
of plaquette-based, or tetramerized, square lattices with no frustration reveal 
a quantum phase transition (QPT) to a plaquette-singlet state at an inter- to 
intra-plaquette coupling ratio $\alpha = 0.55$ \cite{Albuquerque_2008,
Wenzel2008}. A further development of spatial anisotropy and frustration is 
the checkerboard, or ``2D pyrochlore'' lattice \cite{Bishop_2012,Xu_2019_QMC}, 
which also exhibits a QPT to quantum disorder as a function of $J_2/J_1$. In 
the direction of spin anisotropies, Dzyaloshinskii-Moriya (DM) interactions 
\cite{Dzyaloshinskii_1958,Moriya_1960} are the leading consequence of broken 
bond-inversion symmetry in materials, but despite their ubiquity have seen 
rather little attention in square-lattice geometries; available studies 
concern spin ladders \cite{Miyahara2007}, tetramer systems with pyrochlore 
geometry \cite{Kotov2004}, and coupled chains treated by the simplification 
of staggered magnetic fields \cite{Xi2011}. Recent numerical work has explored 
some of the parameter space for frustrated square lattices with exchange 
anisotropies \cite{Rufo2019}. 

Experimentally, the monolayer insulating parent cuprate La$_2$CuO$_4$ has 
been used to obtain accurate measurements of the quantum corrections to 
the spin-wave description of the nearest-neighbor square-lattice model
\cite{Coldea2001}. However, the high energy scales of the cuprate materials 
mean that nontrivial additional physics is involved \cite{Headings2010}, 
possibly including terms beyond quantum magnetism. Of the square-lattice 
compounds with lower energy scales, the most faithful realization is probably 
CFTD \cite{Ronnow_2001}, in which the dynamical properties have been studied 
at all temperatures, while the recent discovery of Sr$_2$CuTeO$_6$ offers 
another candidate with a small $J_2/J_1$ ratio \cite{Babkevich1016}. Beyond 
the nearest-neighbor square lattice, many compounds have been investigated 
as possible realizations of the $J_2/J_1$ model, and while the 
AA$^\prime$VO(PO$_4$)$_2$ vanadium phosphates offer a rich variety 
of (AA$^\prime$) cation options that affect the coupling ratio 
\cite{Tsirlin_2009}, they also suffer from a breaking of 90-degree 
structural symmetry. Perhaps the best realization of the tetramerized square 
lattice is Na$_{1.5}$VOPO$_4$F$_{0.5}$ \cite{Tsirlin_2011}, which opens a 
route towards experimental studes of plaquette-based systems on the frustrated 
square lattice, while La$_2$O$_2$Fe$_2$O(Se,S)$_2$ offers a similar possibility 
for (``double'') checkerboard geometries \cite{Zhu_2010}. However, most studies 
to date have focused on the static properties of these materials, and the 
dynamics of such extended models remain somewhat unexplored.

\begin{figure}[t]
\centering
\includegraphics[width=\linewidth]{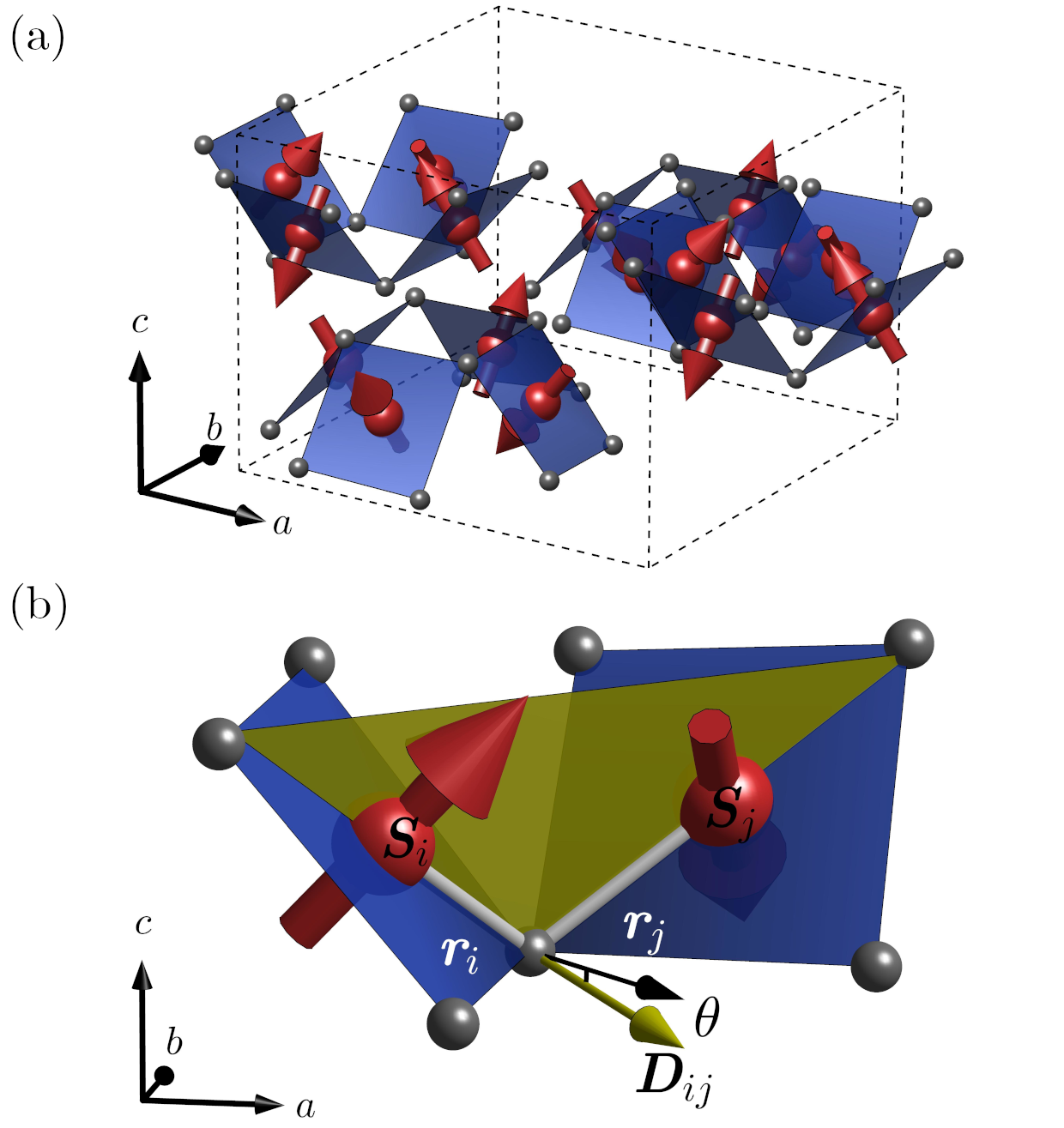}
\caption{\label{fig_magDM} Schematic representation of the ordered magnetic 
structure of Ba(TiO)Cu$_4$(PO$_4$)$_4$. (a) Ordered moments on each Cu atom 
(red) are oriented approximately normal to the CuO$_4$ squares and form a 
two-in, two-out configuration on each cupola, with the relative directions 
between upward- and downward-oriented cupolas as shown. (b) The DM vector 
(yellow) lies in a vertical plane equidistant from the two Cu atoms and 
forms an angle $\theta$ with the horizontal ($ab$) plane.}
\end{figure}

A series of compounds that is known to realize 2D spin-1/2 antiferromagnetism 
on the tetramerized square lattice is the $A$($B$O)Cu$_4$(PO$_4$)$_4$ family, 
where ($A$; $B$) = (Ba, Pb, Sr; Ti) and (K; Nb). Cu$_4$O$_{12}$ tetramers 
form cupola structures, represented in Figs.~\ref{fig:structure} and 
\ref{fig_magDM}(a), which are linked in the $ab$ plane in such a way that 
upward- and downward-oriented cupolas alternate in a checkerboard pattern.  
It was shown using polarized-light microscopy and x-ray diffraction that 
the cupolas also have an alternating rotation about the $c$ axis, shown 
in Figs.~\ref{fig:structure}(b) and \ref{fig:structure}(c), and that the 
extent of this structural chirality depends on the $A^{2+}$ cation 
\cite{Kimura_2016_inorgchem}. This tuneable crystal structure, which 
reaches a highly symmetrical configuration in the (K; Nb) compound 
\cite{Kimura_2020_acs}, was found by a range of thermodynamic measurements 
\cite{Kimura_Ncomms_2016,Kato_2017,Kimura_2018,Islam_2018,Kato_2019,Kumar19,
Kimura_2020_acs} to cause significant changes in the magnetic interactions.  
In fact most of these studies were inspired by the magnetoelectric behavior
that results from ordering of the magnetic quadrupoles formed on the 
Cu$_4$O$_{12}$ tetramers \cite{Kimura_Ncomms_2016,Babkevich_2017}, and also 
leads to nonreciprocal optical properties \cite{Kimura20,Akaki21}. Efforts 
to relate these properties to the structure and geometry of the different 
compounds have to date been based primarily on detailed magnetization 
measurements \cite{Kato_2017,Kimura_2018,Kato_2019,Kimura_2018_Physica_magneto,
Kimura_2019_JPS}. Thus the $A$($B$O)Cu$_4$(PO$_4$)$_4$ family offers a wealth 
of options for exploring how the spin dynamics evolve throughout the 
composition series.

In the present study we begin this investigation by focusing on \sample. It 
was reported by previous powder inelastic neutron studies that the excitation 
spectrum has a robust gap \cite{Kimura_Ncomms_2016}, despite the presence of 
magnetic order, providing an initial hint for the role of DM interactions, 
which are allowed by the rather low symmetry of the Cu-Cu bond pathways in 
this compound. Efforts to extract the magnetic exchange parameters have been 
made on the basis of \textit{ab initio} calculations \cite{Kimura_Ncomms_2016,
Kimura_2018}, also combined with fitting the high-field magnetization response 
for different field directions to a cluster mean field (CMF) approximation 
\cite{Kato_2017,Kato_2019}. These studies suggest a model with dominant 
intra-plaquette interactions, including a strong DM term whose vector 
direction is of key importance, and provide a good description of the 
strong magnetoelectric effect. However, a quantitative benchmarking of 
the proposed interaction parameters by comparison with the spin excitation 
spectrum is precluded by the fact that dynamical measurements have to date 
been possible only with a powder sample \cite{Kimura_Ncomms_2016}. 

We have performed a high-resolution inelastic neutron scattering (INS) 
study of single-crystalline \sample. We observe a complex series of magnetic 
excitations with characteristic periodicities, dispersions, splittings, and 
intensities. From these measurements we determine a definitive set of magnetic 
interaction parameters that describe the dynamical structure factor to high 
accuracy. In contrast to the results based on static quantities, we find that 
the leading inter-plaquette interaction has a magnitude almost identical to 
the leading intra-plaquette one. We obtain five other subsidiary interactions 
with high fidelity and stress the sensitivity of our fits to both in- and 
out-of-plane components of the DM interaction. These results allow us to 
relate \sample~to the ongoing investigation of tetramerized 
``$J_1$-$J_1'$-$J_2$-$J_2'$'' models in extended square-lattice systems.

The manuscript is organized as follows. Section \ref{sec:structure} 
presents the atomic and magnetic structure of Ba(TiO)Cu$_4$(PO$_4$)$_4$. 
In Sec.~\ref{sec:exp_details} we describe the single-crystal growth and 
the experimental procedures used for our INS measurements. Our primary 
results for the dynamic structure factor at zero field are reported in 
Sec.~\ref{sec:results0} and our measurements in applied magnetic fields 
up to 5~T in Sec.~\ref{sec:resultsm}. In Sec.~\ref{sec:resultsfit} we 
present a detailed analysis of our data based on the linear spin-wave 
approximation to the magnetic excitations, from which we deduce the 
optimal set of interaction parameters describing the physics of \sample. 
In Sec.~\ref{sec:mag} we analyze the consequences of these parameters for 
the magnetization and compare these with high-field measurements covering 
the three primary symmetry directions. Section \ref{sd} contains a brief 
discussion connecting the $A$($B$O)Cu$_4$(PO$_4$)$_4$ compounds to the 
general phase diagram of square-lattice models with spatial (tetramerization) 
and spin (DM) anisotropy. A summary and conclusion are provided in 
Sec.~\ref{sec:conclusion}.

\section{Crystal and magnetic structure}
\label{sec:structure}

\sample~has a fascinating and complex crystal structure 
\cite{Kimura_2016_inorgchem}, which we show in detail in 
Fig.~\ref{fig:structure}. As already noted in Sec.~\ref{si}, groups of four 
corner-sharing CuO$_4$ squares form Cu$_4$O$_{12}$ cupola structures, shown 
in blue in Figs.~\ref{fig:structure} and \ref{fig_magDM}. These cupolas 
are connected by PO$_4$ tetrahedra into square-lattice layers in the $(ab)$ 
plane, where both their $c$-axis orientation (up- or down-pointing) and a 
chirality-inducing rotation about the $c$ axis [Fig.~\ref{fig:structure}(c)] 
alternate. Together with the TiO$_5$ pyramids, the PO$_4$ tetrahedra form a 
non-magnetic layer separating the cupola planes [Fig.~\ref{fig:structure}(a)]. 
This tetragonal chiral structure is well described by the $P42_12$ space group, 
with lattice parameters $a = 9.60$ \AA~and $c = 7.12$ \AA. There are 8 
equivalent magnetic atoms in the crystallographic unit cell (Cu, $S = 1/2$), 
whose locations can be generated from original position $(0.27, 0.99, 0.40)$.

Initial studies by magnetic neutron diffraction \cite{Kimura_Ncomms_2016} 
showed the onset of a predominantly antiferromagnetic order below $T_{\rm N}
 = 9.5$ K, with propagation vector $\bm{k} = (0,0,\frac{1}{2})$ and a large 
ordered moment of approximately 0.8$\mu_{B}$. Based on this structural and 
magnetic information, we model the magnetic dynamics of \sample~with the 
Heisenberg Hamiltonian
\begin{equation}
\label{eq:hamiltonian}
\ham = \sum_{[i,j]_m} J_m \, \bm{S}_i \cdot \bm{S}_j - \sum_{\braket{i,j}} \bm{D}_{ij} 
\cdot ( \bm{S}_i \times \bm{S}_j ),
\end{equation}
where $[i,j]_m$ denotes a sum over relevant Cu-Cu bonds with Heisenberg 
interactions of strength $J_m$. Our measurements of the dispersion and 
intensities of the magnetic excitations (Secs.~\ref{sec:results0} and 
\ref{sec:resultsm}) were not previously available, and the aim of our 
analysis is to identify the relevant magnetic interactions, as represented 
in Fig.~\ref{fig:structure}(c), and fit the corresponding $J_m$ values 
(Sec.~\ref{sec:resultsfit}). 

In addition, any understanding of \sample~requires DM interactions, and we 
restrict our considerations to the single term connecting pairs of neighboring 
Cu sites, $\braket{i,j}$, within each cupola. By standard structure and 
symmetry considerations, the DM vectors on the four cupola bonds 
[Fig.~\ref{fig:structure}(c)] are perpendicular to the vector $\bm{r}_i
 - \bm{r}_j$, where $\bm{r}_i$ is the bond vector connecting a Cu site to an 
O atom shared by two CuO$_4$ squares, and are oriented at an angle $\theta$ to 
the $ab$ plane, as shown in Fig.~\ref{fig_magDM}(b). The presence of a large 
DM amplitude, $D = |\bm{D}|$, and the importance of the angle $\theta$ to a 
detailed understanding of the magnetic order, was suggested in the early 
studies of Ref.~\cite{Kimura_Ncomms_2016}. The DM interaction naturally 
frustrates the Heisenberg interactions on each cupola and stabilizes a 
highly non-collinear spin configuration, best understood as a two-in, two-out 
structure [Fig.~\ref{fig_magDM}]. The resulting cupola quadrupole moment has 
been characterized in detail by spherical neutron polarimetry (SNP) 
\cite{Babkevich_2017} and nuclear magnetic resonance (NMR) measurements 
\cite{Rasta_2020}. 

\section{Material and methods}
\label{sec:exp_details}

Three single crystals of \sample~with a total mass of 3.3~g were grown by the 
flux method \cite{Kimura_2016_inorgchem}. They were co-aligned on an Al holder 
to a precision of less than 1$^\circ$ in the $(hk0)$ scattering plane using Laue 
x-ray diffractometry. Initial measurements of the spin dynamics at zero 
magnetic field were performed on the MACS spectrometer \cite{Rodriguez08}
at NIST, and revealed a complex spectrum of modes that could not be 
resolved completely in parts of the Brillouin zone. To achieve a better 
resolution in the required energy ranges, further experiments were performed 
at zero field on the direct-geometry time-of-flight (ToF) neutron spectrometer 
IN5 \cite{DOI_IN5}, and in a vertical field applied along the sample $c$ axis 
on the triple-axis spectrometer (TAS) IN12 \cite{DOI_IN12}, both at the 
Institut Laue-Langevin (ILL). 

\begin{figure}[p]
\centering
\includegraphics[width=\linewidth]{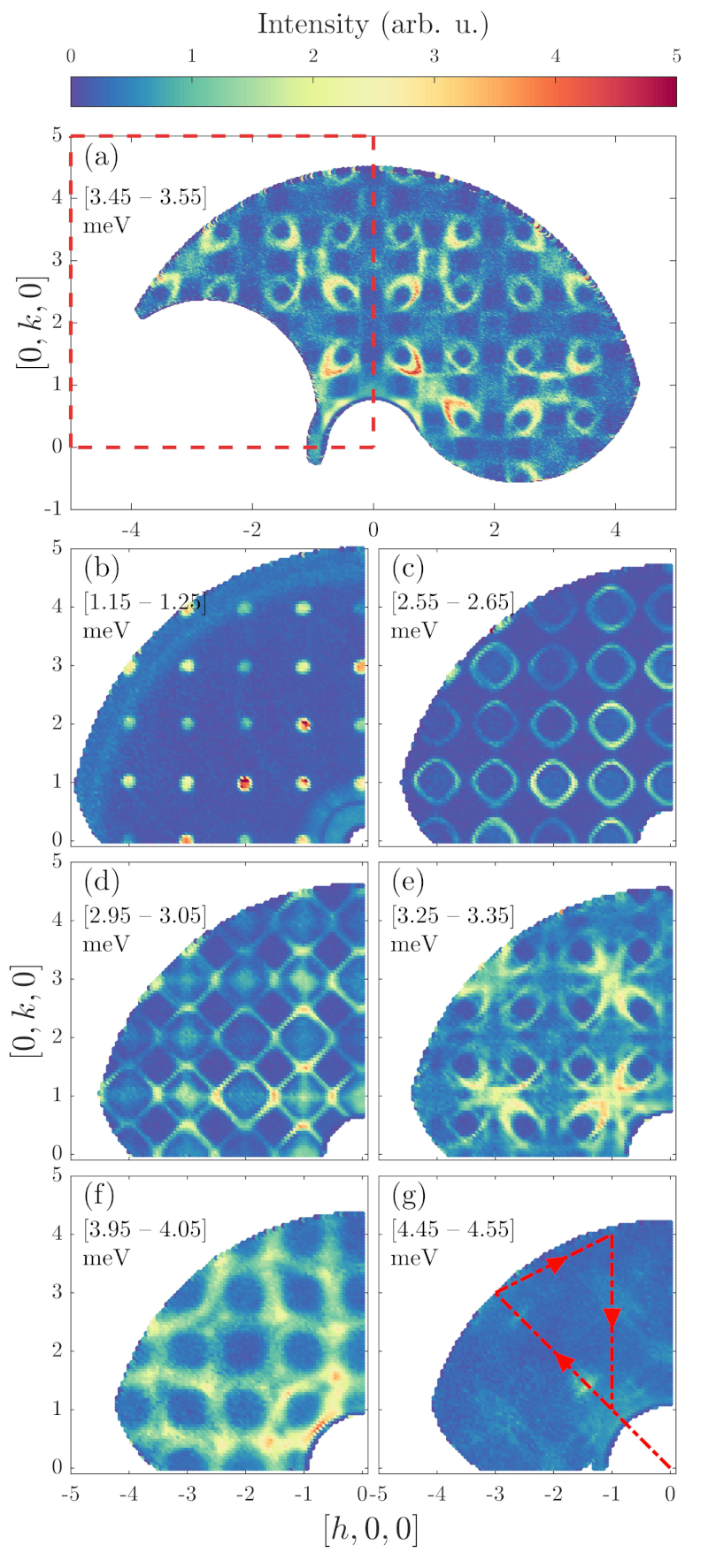}
\caption{\label{fig:cst_E} Scattering intensity, $I(\bm{Q},\omega)$, 
integrated over different $\omega$ ranges throughout the band width of the 
excitations and shown as a function of $\bm{Q}$ in the ($hk0$) scattering 
plane. The measurement temperature was 2~K, the momentum step was $dq = 0.01$ 
\AA$^{-1}$, and the energy integration range is indicated in each panel. (a) 
Unsymmetrized data across the full accessible Brillouin zone, highlighting 
the fourfold symmetry of the excitations in $\bm{Q}$. The dashed red box 
marks the second quadrant, to which the remaining panels should be referred. 
(b)-(g) Symmetrized data, folded onto the second quadrant, and shown for 
six selected energy ranges. Red dashed lines in panel (g) show the scattering 
wave vectors presented in Fig.~\ref{fig:spinw_fit}.} 
\end{figure}

On IN5, measurements were made at 1.5~K in the ordered phase, at 10~K just 
above $T_{\rm N}$, and deep in the paramagnetic phase at 30~K. Inelastic data 
were collected by rotating the sample around its $c$ axis by a total of 
138$^{\circ}$, in steps of 1$^{\circ}$. Counting times were 20 minutes per 
angular step at 1.5~K and 30~K, and 13 minutes per step at 10~K, for a total 
measurement time of 46 hours. The crystals were oriented in order to maximize 
the accessible range in the $(hk0)$ plane, and such that scattering in the 
orthogonal direction could be measured using the opening of the orange cryostat.
The incident energy was set to $E_{i} = 7.08$ meV and the chopper rotation speed
to 200 Hz, resulting in resolutions of 0.24 meV (FWHM) at the elastic line, 
decreasing to 0.15 meV (FWHM) at the highest energy transfers. We took 
advantage of the tetragonal symmetry of \sample~by summing the intensities 
from detection pixels corresponding to $\bm{Q}$ points that are equivalent 
under crystal symmetry operations of the point group (422). Due to the 
non-dispersive behavior of the excitations along [$0,0,l$], the data were 
integrated over $\pm 0.6$ in $l$. The ToF data were processed using the 
Horace software suite \cite{Ewings_2016}.

On IN12, measurements were made at a base temperature of 2~K. The final wave 
vector was fixed to $k_f = 1.3 \text{~\AA}^{-1}$, giving a resolution of 0.172(5)
meV (FWHM). An 80$^\prime$ collimator was placed between the monochromator 
and the sample; the monochromator had both horizontal and vertical focusing 
while the analyzer was horizontally focused only. The sample was inserted in 
a 10~T vertical cryomagnet, in which data were collected at field values up 
to 5~T in 1~T steps. The counting time was 2 minutes per $\bm{Q}$-point. In 
both experiments, the intensity $I(\bm{Q},\omega)$ measured at each wave-vector 
transfer, $\bm{Q}$, and energy transfer, $\omega$, is directly proportional to 
the dynamical structure factor, $S(\bm{Q},\omega)$, convolved with a Gaussian 
distribution to account for the finite measurement resolution of each 
spectrometer.

\begin{figure}[t]
\centering
\includegraphics[width=0.9\linewidth]{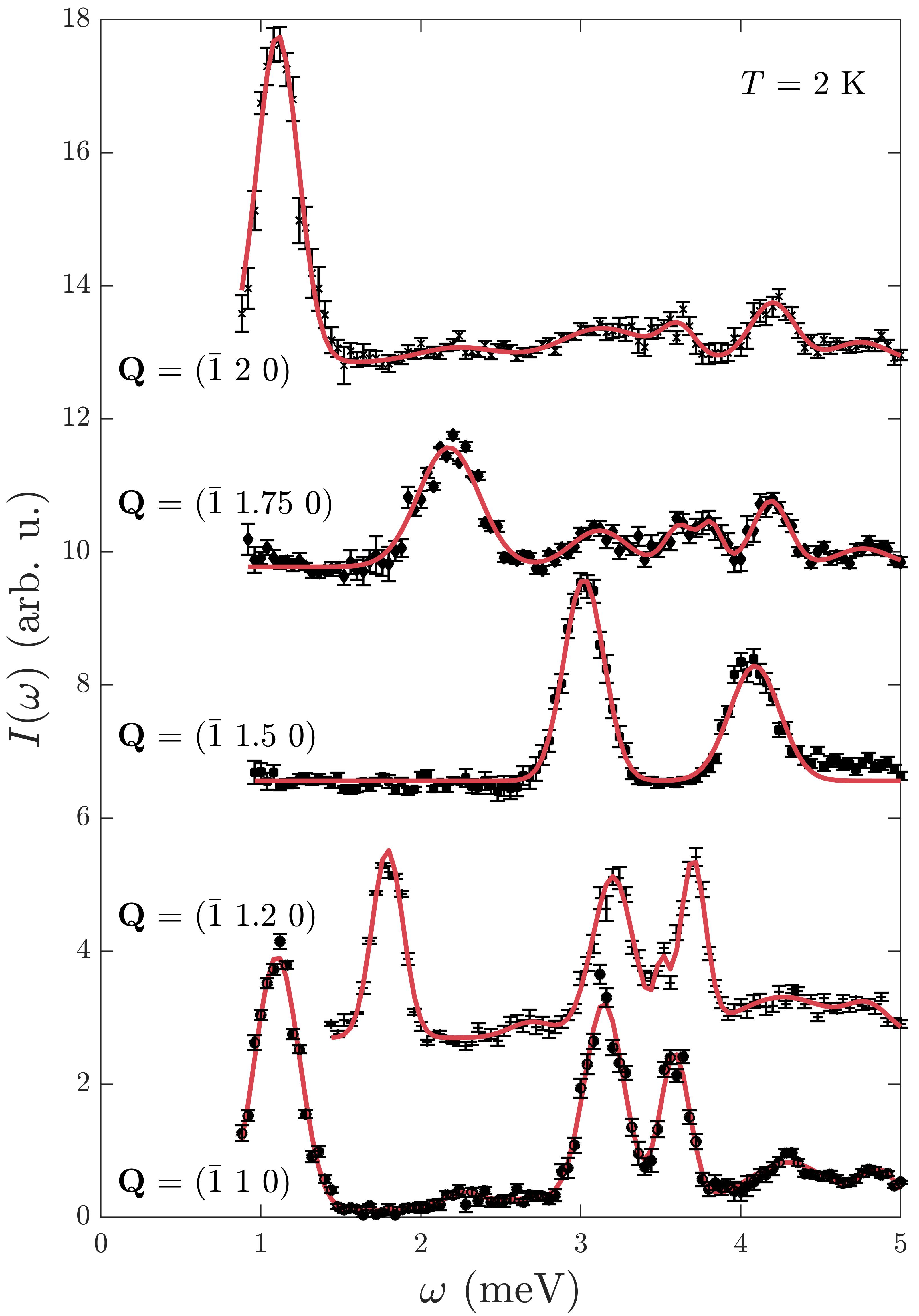}
\caption{\label{fig:TOF_cuts} $I(\bm{Q},\omega)$ (black points) collected
at 2 K for different $\bm{Q}$ points along the [$\bar{1},k,0$] direction
and shown as a function of $\omega$. The red lines are an interpolated 
multi-Gaussian fit, from which the peak centers and widths were extracted.}
\end{figure}

\begin{figure*}[t]
\centering
\includegraphics[width=\textwidth]{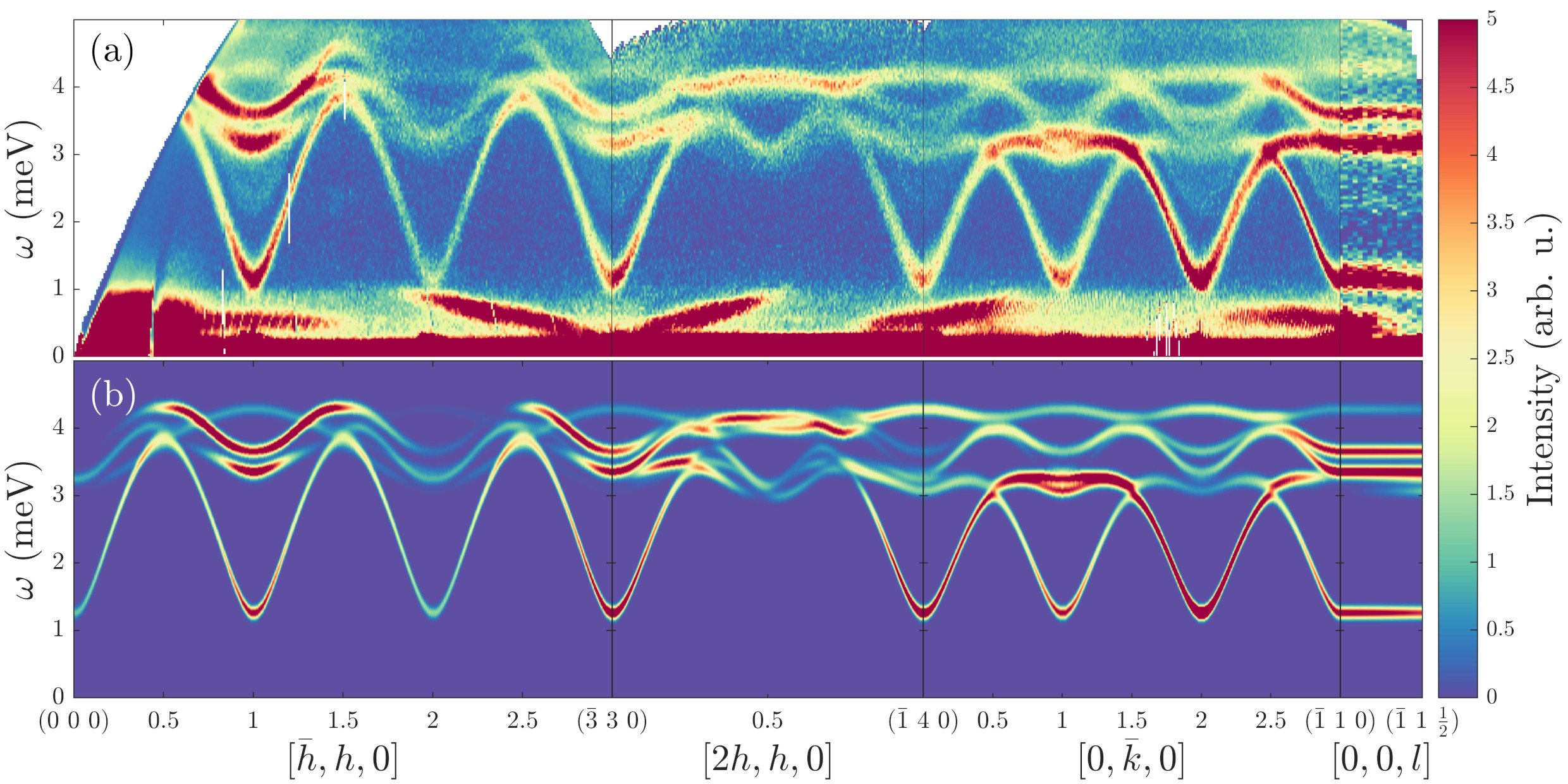}
\caption{\label{fig:spinw_fit} (a) Measured dynamical structure factor, 
$S(\bm{Q},\omega)$, shown for the three high-symmetry $\bm{Q}$ directions
indicated in Fig.~\ref{fig:cst_E}(g) and for the out-of-plane direction. 
The step size in energy is $dE = 0.04$ meV and in momentum it is $dq = 0.01$ 
\AA$^{-1}$ in the scattering direction for $\bm{Q}$ in the ($hk0$) plane and 
$dq = 0.03$ \AA$^{-1}$ for $\bm{Q}$ in the [$0,0,l$] direction. The integration 
range in the orthogonal direction is $\pm 0.06$ \AA$^{-1}$. No smoothing effects 
were introduced to present these data. (b) $S(\bm{Q},\omega)$ modelled using a 
spin-wave description of the dispersion convolved with Gaussian functions 
representing the spectrometer resolution.}
\end{figure*}

\section{Spin dynamics}

\subsection{Zero applied field}
\label{sec:results0}

We begin by reporting the INS spectrum of \sample~at zero magnetic field, 
as measured on IN5. Figure~\ref{fig:cst_E} presents $I(\bm{Q},\omega)$ at 
2 K as color maps of the magnetic excitations integrated over selected 
constant energy ranges at $\bm{Q}$ values throughout the Brillouin zone (BZ). 
This format confirms that the spectrum has the same fourfold symmetry as the 
atomic structure [Fig.~\ref{fig:cst_E}(a)], which justifies averaging the 
measured intensity and discussing a single quadrant in the remaining panels. 
As the energy is increased, Fig.~\ref{fig:cst_E}(b) 
shows that gapped spin excitations appear first at the Bragg-peak positions 
at an energy of 1.13(3) meV. This branch shows a strong dispersion for 
wave vectors across the BZ [Fig.~\ref{fig:cst_E}(c)], and at approximately 
3 meV they begin to merge while a different excitation branch also emerges at 
the Bragg peaks [Fig.~\ref{fig:cst_E}(d)]. In the energy range up to 4 meV, 
scattering contributions from several different branches disperse and merge, 
resulting in complicated patterns in $\bm{Q}$ [Figs.~\ref{fig:cst_E}(e) and 
\ref{fig:cst_E}(f)], but ones that always retain the same periodicity. Finally, 
above 4.4 meV one finds only weak remnant scattering [Fig.~\ref{fig:cst_E}(g)]. 

To study the evolution of these spin excitations with 
$\omega$, Fig.~\ref{fig:TOF_cuts} shows representative constant-$\bm{Q}$ scans 
taken along a single high-symmetry direction in reciprocal space. We observe 
the presence of multiple sharp peaks, all of which are well described by a 
Gaussian line shape. By extracting peak centers, widths, and intensities in 
this way, we identify a maximum of seven different excitations in some 
parts of the BZ. 

Figure \ref{fig:spinw_fit}(a) collects this information to display the 
dynamical structure factor, $S(\bm{Q},\omega)$, along several different 
high-symmetry $\bm{Q}$-space directions. The spectrum is extremely rich, 
and to describe it we begin by decomposing the observed excitations into 
three distinct regimes of energy, which we define on the basis of the 
[$0,\bar{k},0$] scattering direction (the third panel in 
Fig.~\ref{fig:spinw_fit}). First, there is a robust gap, $\Delta = 1.13(3)$ 
meV, at the BZ center, and in fact this repeats along all measured directions.
Second, a single, sharp excitation branch with a largely cosinusoidal 
dispersion is present at 1--3 meV, to which we refer henceforth as the 
low-energy regime. At the BZ boundaries, this mode flattens in a manner 
reminscent of a level-repulsion with the higher-energy excitations. The 
gradient with which the low-energy mode disperses around the Bragg-peak 
positions indicates the magnitude of the leading interaction, and the fact 
that this mode seems to have a periodicity of two BZs (Fig.~\ref{fig:cst_E}) 
suggests that this interaction spans half of the magnetic unit cell. Third, 
the high-energy regime at 3--4.5 meV contains three distinct and continuous 
modes, one of which merges into the low-energy mode at the lower edge of the 
energy window. We comment again that there are no magnetic excitations above 
the upper edge of the high-energy regime [Fig.~\ref{fig:cst_E}(g)].

We also report a number of subtle details in Fig.~\ref{fig:spinw_fit}(a), 
which are important for different aspects of fitting the relevant interaction 
parameters. Above the low-energy mode one may discern the presence of an 
additional scattering feature with very low intensity; denoting the low-energy 
mode dispersion by $E_1(\bm{Q})$, this feature appears above $E_\text{2M}(\bm{Q})
 = E_1(\bm{Q}) + \Delta$. This information allows us to identify the feature as 
a two-magnon scattering continuum, which is sharpest at its lower boundary, and 
in Sec.~\ref{sec:resultsm} we will obtain further information to confirm this 
identification. Another important detail is the splitting of the second most 
energetic mode that we observe around the zone centers, as this is a 
consequence of the DM interactions and, together with the gap, provides the  
most accurate means of quantifying $D$; this feature, at 3.2--3.5 meV, is 
clearest for the [$0,\bar{k},0$] direction. In general, the scattering 
intensity is strongest near the zone centers, and in the low-energy regime, 
although a clear exception occurs in the second BZ, where the high-energy 
branches are equally intense. We also observe an expected drop in scattering 
intensity with increasing $\bm{Q}$ that arises from the magnetic form factor 
of Cu. As expected from the crystallographic structure, the magnetic 
excitations are only very weakly dispersive in the direction orthogonal to 
the magnetic layers [right panel of Fig.~\ref{fig:spinw_fit}(a)], and thus 
we have chosen to integrate all of our scattered intensities over the range 
$-0.6 < l < 0.6$.

\subsection{Vertical magnetic field}
\label{sec:resultsm}

We turn now to the evolution of the spin excitation spectrum of \sample~in 
the presence of an applied magnetic field. The field adds a term 
\begin{equation}
\ham_m = - g \mu_B \sum_i \bm{B} \cdot \bm{S}_i\,,
\label{eq:ham_zeeman}	
\end{equation}
to the Hamiltonian of Eq.~(\ref{eq:hamiltonian}), where the $g$-factor 
is assumed to be isotropic and $\mu_B$ is the Bohr magneton. The sample 
was aligned on IN12 such that the field was applied along the $c$ axis, 
i.e.~$\bm{B} = (0, 0, B_z)$. The vertical magnetic field provides significant 
insight not only into the degeneracy of the $B = 0$ magnetic excitations, 
which in general should split in the presence of $B_z$, but also into the 
effects of the DM interaction, because $[\ham_m, \ham_D] \neq 0$, where 
$\ham_D$ is the second term of Eq.~(\ref{eq:hamiltonian}), in any situation 
other than $\bm{D}_{ij} = D_z$ ($\theta = 90^{\circ}$). 

\begin{figure}[t]
\centering
\includegraphics[width=0.95\linewidth]{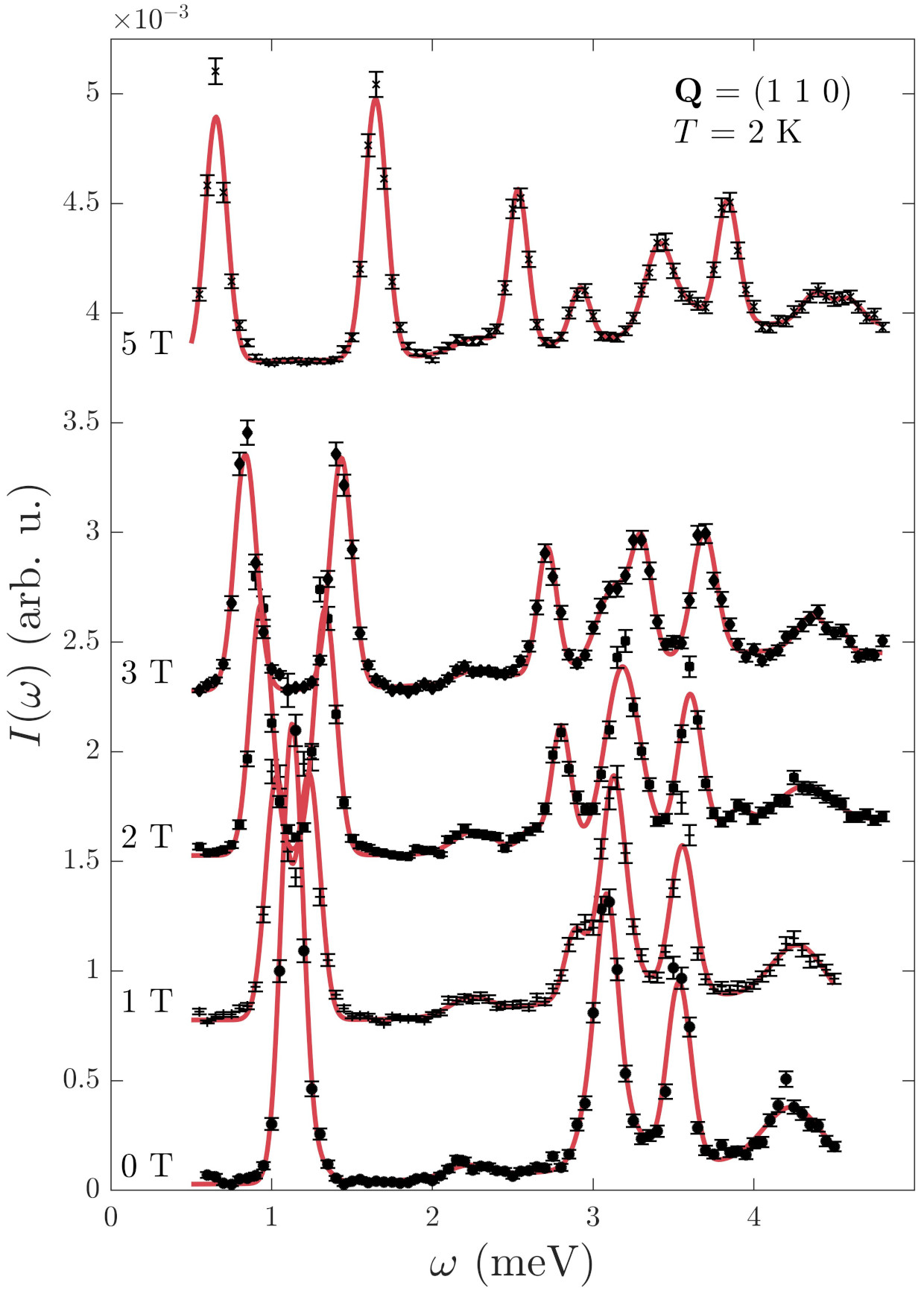}	
\caption{\label{fig:IN12_field_cuts} $I(\bm{Q},\omega)$ measured at $\bm{Q}
 = (1 \; 1 \; 0)$ in vertical magnetic fields of 0, 1, 2, 3, and 5 T. Black 
points denote measured intensities, normalized by monitor counts, and red 
lines are obtained from a fit to multiple Gaussian functions.}	
\end{figure}

Although one may anticipate from the size of the gap and band center 
(Sec.~\ref{sec:results0}) that small fields have little effect on the 
excitation spectrum, the instrumental resolution of IN12 allowed us to 
distinguish the split modes even at 1 T. Figure \ref{fig:IN12_field_cuts} 
shows the measured scattering intensities, represented as energy scans at a 
constant $\bm{Q} = (1 \; 1 \; 0)$, for fields of 0, 1, 2, 3, and 5 T. The 
three sharp modes of the zero-field spectrum shift and split progressively, 
until at 5 T we observe six well-resolved peaks, all of which are well 
described by Gaussian profiles. The lowest mode splits clearly into two 
branches, of equal scattering intensity, which move symmetrically down and 
up in energy with increasing field. By contrast, the energies and intensities 
of the two modes in the high-energy regime show a more complex evolution, on 
which we comment below. 

The spectra of Fig.~\ref{fig:IN12_field_cuts} contain two additional features. 
One is a broad hump of scattering intensity above 4 meV, which appears to 
move upwards with field until at 5 T it is centered at 4.5 meV. The other is 
a broad and weak excitation around 2.1~meV, which we identified previously 
as a continuum of two-magnon scattering processes. This feature does not move 
as the field is increased, which is consistent with processes creating two 
spin waves of $\Delta S^z = 1$ and $-1$, such that the composite $\Delta 
S^z_{\text{tot}} = 0$ excitation does not respond to an external magnetic field. 

\begin{figure*}[t]
\centering
\includegraphics[width=0.94\linewidth]{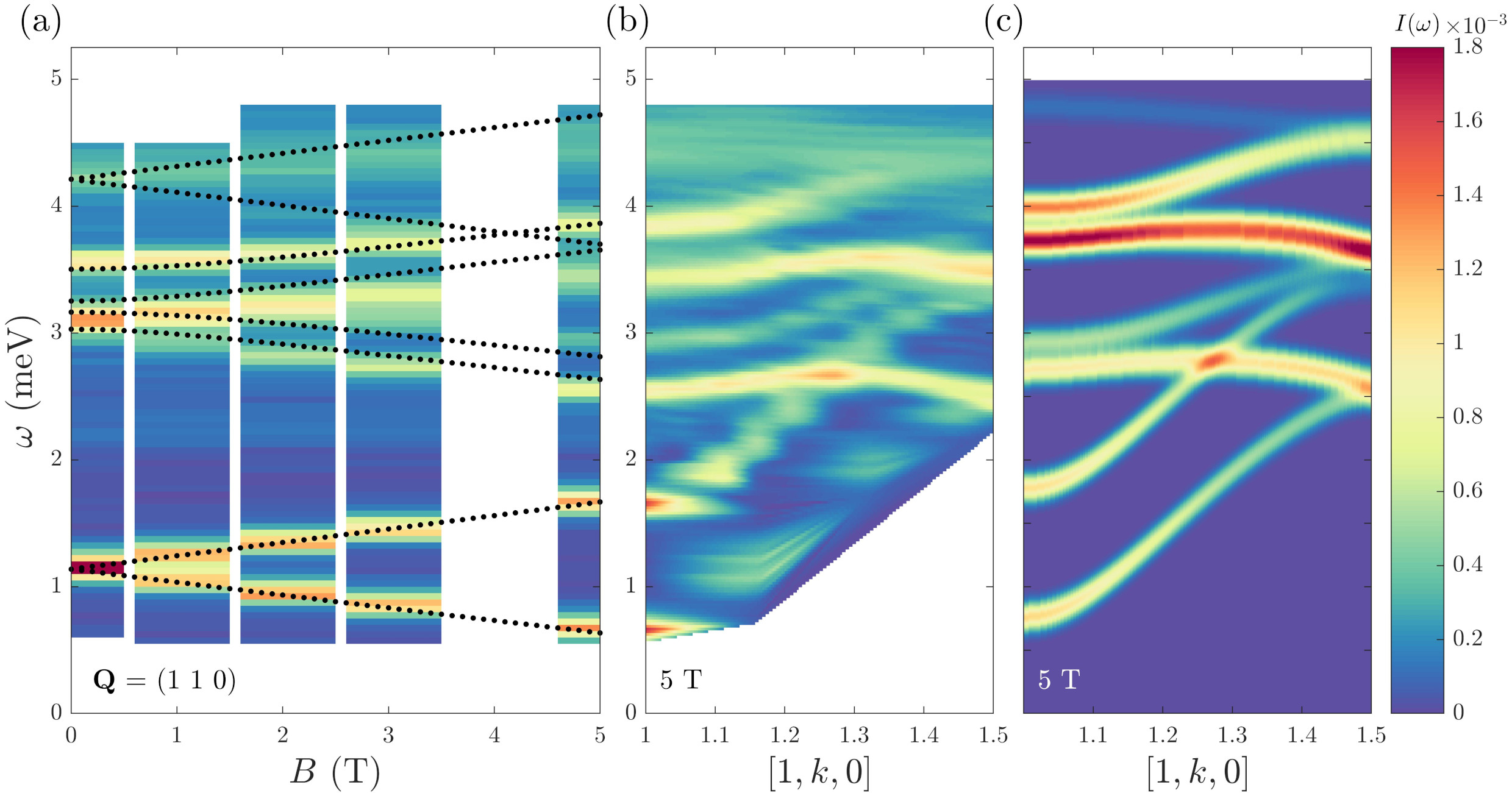}
\caption{(a) Intensity data of Fig.~\ref{fig:IN12_field_cuts} presented as 
a color map for the five measured magnetic field values. Black points show 
the energies of the magnetic excitations at $\bm{Q} = (1 \; 1 \; 0)$ obtained 
from LSW calculations based on the Hamiltonian of Eq.~(\ref{eq:ham_zeeman}) 
with the parameters of Table~\ref{tab_Jparam}. (b) Excitation spectrum measured 
at 5 T along the $[1,k,0]$ direction. For a better representation of 
these data, a linear interpolation was applied to measurements at discrete 
$\bm{Q}$ points of the type shown in Fig.~\ref{fig:IN12_field_cuts}. (c) 
Excitation spectrum in a field of 5~T, modelled in the LSW approximation for 
this half-BZ $\bm{Q}$-scan.}
\label{fig:sw_field_dispersion} 
\end{figure*}

The field-induced evolution of the spin-wave branches fitted by Gaussians 
in Fig.~\ref{fig:IN12_field_cuts} is represented as a color map in 
Fig.~\ref{fig:sw_field_dispersion}(a). The low-energy mode and particularly 
the broad peak above 4 meV show a near-ideal linear splitting from very low 
fields, whereas the peaks in the 3--3.6 meV regime at $B = 0$ remain rather 
flat for $B \lesssim 0.5$~T before recovering the same gradient beyond 1.5~T. 
This indicates differing degrees of sensitivity to the in-plane (non-commuting) 
component of the DM interaction, although we note that the complex geometry 
of these interactions on each plaquette [Fig.~\ref{fig:structure}(c)] makes 
it difficult to equate the field scale with $D$. The gradients of the linear 
(Zeeman) evolution beyond 2 T match for all of the split branches observed, 
with the peak centers of the lowest mode, moving by $-0.09(3)$ meV/T and $0.
10(3)$ meV/T. These slopes are consistent with the value $g \mu_{\rm{B}} = 0.12$ 
meV/T expected for a spin-1 excitation. The field-induced behavior shown in 
Fig.~\ref{fig:sw_field_dispersion}(a) allows us to deduce the origin of the 
peaks in the intermediate-energy regime. Of the two peaks apparent at $B = 0$, 
the one around 3 meV in fact contains three branches, while the one at 3.5 meV 
is a single branch. These modes are not degenerate at $B = 0$ because of the 
DM interaction, which generates the 0.5 meV separation of the $\Delta S^z = 
\pm1$ branches. 

We have extracted the dynamical structure factor at 5 T from several 
$\omega$-scans, of the type shown in Fig.~\ref{fig:IN12_field_cuts}, measured 
at multiple $\bm{Q}$ points, and in Fig.~\ref{fig:sw_field_dispersion}(b) we 
show the excitation spectrum over half of the BZ in the $[1,k,0]$ direction. 
These results verify that each set of split modes disperses in the same way 
with $\bm{Q}$, i.e.~that the effects of the field are the same on each branch 
at each $\bm Q$. No additional splittings are observed with $\bm{Q}$, indicating
that $\bm{Q} = (1 \; 1 \; 0)$ has no special symmetries. The different branches 
disperse differently and we comment that at 5 T they simply cross, showing 
no evidence of the avoided crossings associated with level mixing. 

\subsection{Magnetic Hamiltonian}
\label{sec:resultsfit}

We now propose a set of parameters that, when inserted in the spin Hamiltonian 
of Eq.~(\ref{eq:hamiltonian}), describes the dynamical structure factor of the 
system both in dispersion and in intensity. In a material with the structure 
of \sample~[Fig.~\ref{fig:structure}(c)], there are {\it a priori} two 
separate possibilities for opening the observed gap, the tetramerization 
and the DM interaction. Motivated by the relatively large band width of 
the low-energy excitation [Fig.~\ref{fig:spinw_fit}(a)] and the interaction 
parameters estimated from static measurements \cite{Kato_2017}, we adopt the 
hypothesis that the system is not strongly tetramerized and the gap arises 
primarily from the DM term. In addition, the robust ordered moment of 
\sample~suggests that a linear spin-wave (LSW) theory should provide a good 
approximation in which to describe the magnetic order and excitations, and 
thus we employ the SpinW package \cite{spinw}. 

The optimal set of parameters required in Eq.~(\ref{eq:hamiltonian}) is 
obtained by fitting the measured dispersions throughout the BZ at zero field 
[Fig.~\ref{fig:spinw_fit}(a)], with additional information taken from the 
available finite-field dispersions [Fig.~\ref{fig:sw_field_dispersion}]. The 
geometry of these interactions is shown in Fig.~\ref{fig:structure}(c) and 
their values are given in Table \ref{tab_Jparam}. A quantitative estimate of 
the uncertainties on the strongly interdependent Heisenberg parameters is 
difficult to extract from SpinW; by contrast, the uncertainty in $D = 1.07(3)$ 
is established quite directly by the measured spin gap. The ordered ground 
state corresponding to these parameters is qualitatively similar to that 
deduced from SNP measurements \cite{Babkevich_2017} and shown in 
Fig.~\ref{fig_magDM}, with the Cu spins forming a canted two-in, two-out 
arrangement on each cupola. However, the spin direction in the zero-field 
ground state estimated in the SNP analysis is almost normal to the CuO$_4$ 
squares, making an angle of $45^{\circ}$ with the ($ab$) plane, whereas for 
the ground state deduced from the LSW description this angle is $63^{\circ}$. 
We return to this issue after discussing the relative values of $D$ and the 
Heisenberg parameters $\{J_m\}$. 

The zero-field excitation spectrum produced with the model parameters of Table 
\ref{tab_Jparam} is shown in Fig.~\ref{fig:spinw_fit}(b). It is clear that all 
the primary features of the measured bands are captured with quantitative 
accuracy. Crucial confirmation of this parameter set is provided by the fact 
that the scattered intensities are very well reproduced with no further 
fitting. The level of the remaining discrepancies is extremely small, and 
concerns mostly details of apparent (anti-)crossing events between rather flat 
modes in the high-energy regime, although some of these may be a consequence 
only of low intensities. We note that the feature $E_\text{2M} (\bm{Q})$ with 
onset around 2.1 meV is not present in the fitted spectrum, consistent with our 
conclusion that it is not an elementary spin wave but a two-magnon scattering 
state. Concerning the field-induced evolution of these modes, again the fits 
(dotted lines) in Fig.~\ref{fig:sw_field_dispersion}(a) show only very minor 
deviations from the measured data for only one of the multiplets at 
intermediate energies.

\begin{table}[b]
\caption{\label{tab_Jparam}
Interaction parameters, in meV, used in the LSW description of the magnetic 
spectrum of Figs.~\ref{fig:spinw_fit}(a) and \ref{fig:sw_field_dispersion}(b). 
The geometry of these interactions is shown in Fig.~\ref{fig:structure}(c) and 
the meaning of the angle $\theta$ in Fig.~\ref{fig_magDM}(b).}
\begin{ruledtabular}
\begin{tabular}{cccccccc}
$J_1$ & $J_2$ & $J_{2}'$ & $J_{11}'$ & $J_{12}'$ & $D$ & $\theta$
\\ \colrule
2.03 & 0.52 & 2.22 & 0.17 & 0.17 & 1.07 & 10$^{\circ}$ \\
\end{tabular}
\end{ruledtabular}
\end{table}

The Heisenberg interactions of Table \ref{tab_Jparam} define a magnetic 
lattice of square antiferromagnetic plaquettes, $J_1$, with a small diagonal 
intra-plaquette coupling, $J_2$, generating rather weak frustration. The 
dominant interaction linking the plaquettes in the $ab$ plane is not $J_1'$, 
the bond that would form a conventional square lattice, but the diagonal 
coupling, $J_2'$. This result is consistent with the geometry of the 
Cu--O--P--O--Cu bonds connecting the plaquettes, which, as represented in 
Fig.~\ref{fig:structure}(b), almost form a single curve for $J_2'$ but include 
an additional 90$^\circ$ kink for $J_1'$. The magnitude of $J_2'$ can be fitted 
to high accuracy from the disperion of the lowest mode, and one of our most 
striking results is that the optimal $J_2'$ is slightly (10\%) larger than 
$J_1$ (whose Cu--O--Cu bond angle is only 108$^\circ$). This implies that the 
degree of tetramerization contained within the Heisenberg parameters alone 
is rather small. It also indicates that the inter-plaquette coupling is 
twice as strong as the value proposed in Ref.~\cite{Kato_2017}, and in 
Sec.~\ref{sec:mag} we investigate this discrepancy. 
Finally, the structure of \sample~requires two different $J_1'$ bonds, which 
we label $J_{11}'$ and $J_{12}'$, and we find these to be similar in value but 
weak by comparison with $J_2'$ (Table \ref{tab_Jparam}). In the spin-wave 
spectrum of Fig.~\ref{fig:spinw_fit}, these interactions are necessary for 
an accurate description of the separation between closely-spaced modes in 
the high-energy regime, particularly around 3 meV, and our fitting quality 
deteriorates when they are not equal and antiferromagnetic. By contrast, 
these two parameters were given opposite signs in fitting the magnetization 
data, suggesting that smaller parameters in the global fit can be subject to 
large relative uncertainties. 

The other strong interaction in Table \ref{tab_Jparam} is the DM term, whose 
vector nature results in two unknown parameters, equivalently $(D, \theta)$ 
or the projections $D_\| = D \cos \theta = 1.05(3)$ meV in the $ab$ plane 
(orthogonal to the $J_1$ bond) and $D_z = D \sin \theta = 0.18$ meV along the 
$c$ axis. By symmetry, the vector $\bm{D}$ lies in the plane orthogonal to 
the Cu--Cu bond of $J_1$ and its direction alternates between all-in or all-out 
[Fig.~\ref{fig:structure}(c)] with the upward or downward cupola orientation. 
The strong $J_2'$ interaction means that the origin of the gap must lie in the 
DM term, and thus it is no surprise to find a large magnitude, $D$. In more 
detail, the fitted gap is extremely sensitive to the value of $D_\|$, fixing 
its value within a narrow window, whereas any value of $D_z$ below $0.8$~meV 
has rather little effect. 

For a more accurate determination of the direction, $\theta$, of the DM 
vector we exploit the fact that its in- and out-of-plane components have quite 
different effects on the SU(2)-symmetric eigenstates of the Heisenberg terms 
in $\ham$ [Eq.~(\ref{eq:hamiltonian})] and on the Zeeman-split eigenstates in 
the presence of in $\ham_m$ [Eq.~(\ref{eq:ham_zeeman})]. In zero field it is 
easy to demonstrate that the parameters of Table \ref{tab_Jparam} provide a 
consistent description of certain mode separations in the high-energy regime, 
which cannot be achieved using the inter-plaquette $J_m$ parameters alone, but 
it is difficult to demonstrate uniqueness. By contrast, the finite-field data 
we show in Fig.~\ref{fig:sw_field_dispersion}(a) provide detailed information 
about field gradients and anti-crossings that are reproduced accurately 
(black dotted lines). The 5 T dispersion and intensity data of 
Fig.~\ref{fig:sw_field_dispersion}(b) are also fitted with quantitative 
accuracy by the LSW description with these $\bm{D}$ parameters, as shown in 
Fig.~\ref{fig:sw_field_dispersion}(c). These LSW calculations also indicate 
that $D_z$ values in excess of 0.18 meV cause a visible splitting of the 
2.5~meV mode in Fig.~\ref{fig:sw_field_dispersion}(c), which sets an upper 
limit on this quantity. 

The resulting value, $\theta = 10^{\circ}$, is in complete agreement with the 
conclusions drawn from static measurements \cite{Kato_2017}. On structural 
grounds one might expect this angle to take the value $\theta_p = 14^{\circ}$ 
obtained for a single cupola bond from $\bm{D}_{ij} \propto \bm{r}_i \times 
\bm{r}_j$ [Fig.~\ref{fig_magDM}(b)]. Thus both static and dynamic measurements 
indicate only minimal corrections to this expectation, despite the potentially 
complex spin-density distribution in the full cupola wave function. Returning 
to the spin orientation in the ground state, the LSW result that the ordered 
moments on the square are canted at 27$^\circ$ from the $c$ axis translates 
to an angle of 37$^\circ$ between the two anti-aligned spins on a bond. Given 
the relative strengths of the leading $J$ terms, which favor collinear order, 
and the DM term, which favors a 90$^\circ$ angle, the value of 37$^\circ$ is 
fully consistent. 

As noted in Sec.~\ref{si}, the greater data volume provided by the dynamical 
excitations and their higher sensitivity to the coupling parameters of the 
system allows us to obtain a more accurate account of the magnetic interactions 
than was possible using static measurements. The primary point of 
difference with the previous results \cite{Kimura_Ncomms_2016,Kato_2017} is 
the much larger inter-plaquette coupling provided by $J_2'$. Next we provide  
(Sec.~\ref{sec:mag}) a more detailed discussion of the implications of this 
result for fitting the high-field magnetization data and for understanding 
the further properties of the system, including magnetoelectricity. 

Before turning to this issue, we close our discussion of parameters by noting 
that a weak interlayer interaction, $J_{\perp}$, is required to ensure the 
observed antiferromagnetic order. In principle this parameter could be fitted 
from the very weakly dispersive behavior of the low-energy mode for wave 
vectors $\bm{Q}$ in the $[0,0,l]$ direction (Fig.~\ref{fig:spinw_fit}). In 
the present experiment, geometrical and resolution factors were such that our 
data for the out-of-plane direction are of qualitative value only, and thus 
we did not attempt to include $J_{\perp}$ in our fitting procedure. All of our 
observations are consistent with the order-of-magnitude estimate $J_{\perp} 
\approx J_1/100$ proposed in previous studies \cite{Kato_2017}. The resulting 
strongly 2D nature of the $A$($B$O)Cu$_4$(PO$_4$)$_4$ family of compounds, 
combined with the clearly resolvable effects of all the different parameters 
in Table \ref{tab_Jparam}, makes them valuable candidates for investigating 
quantum phases in spatially and spin-anisotropic square-lattice models, as 
we discuss further in Sec.~\ref{sd}.

\begin{figure*}[t]
\centering
\includegraphics[width=0.92\linewidth]{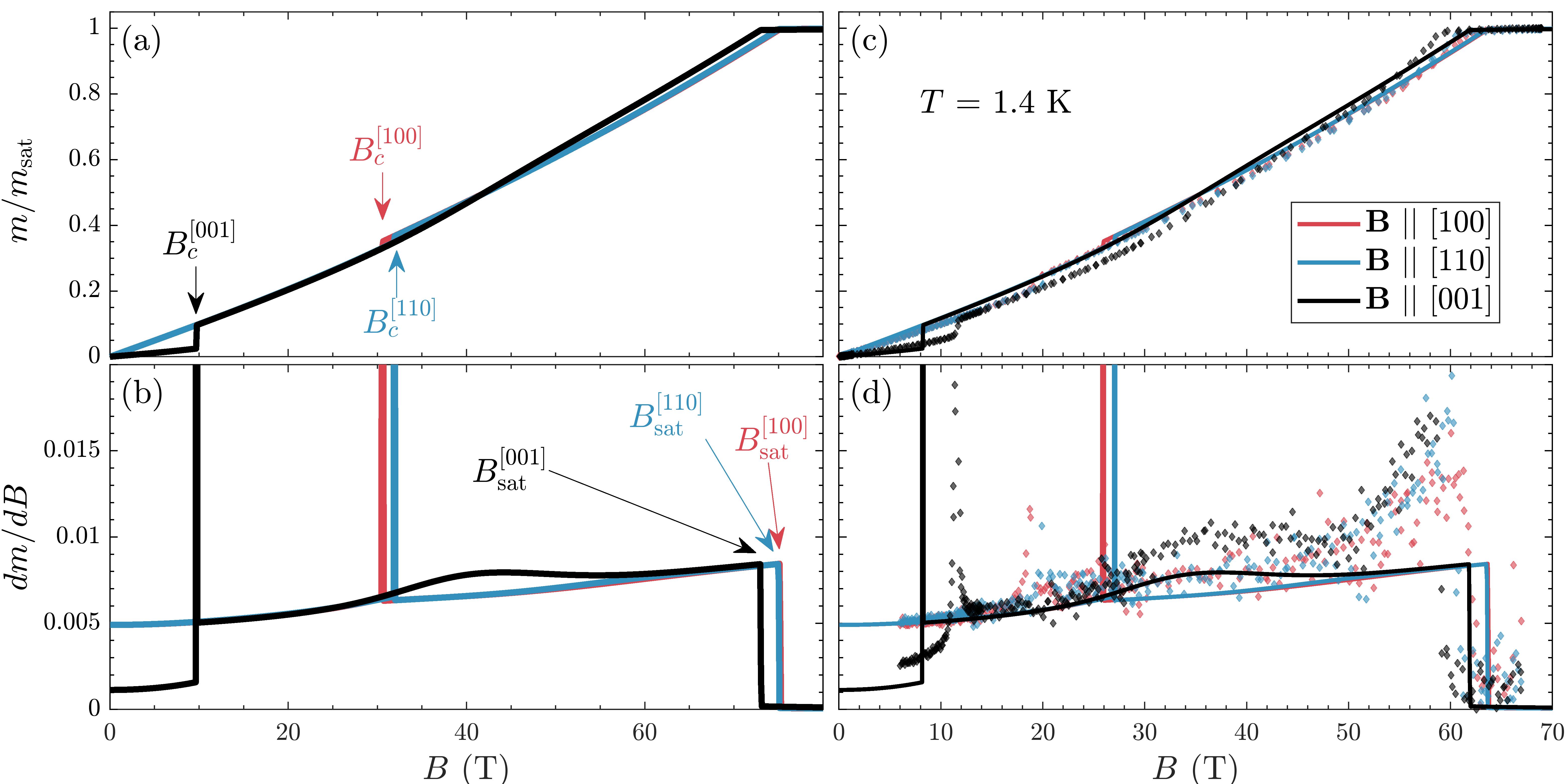}
\caption{(a) Magnetization, $m(B)$, calculated using the magnetic 
interaction parameters of Table \ref{tab_Jparam}, for fields applied in 
the three primary crystallographic directions of the tetragonal structure 
(Fig.~\ref{fig:structure}). $m_{\rm sat}$ denotes the saturation magnetization 
and is used for normalization. (b) Calculated magnetization derivative, 
$dm(B)/dB$, which highlights the discontinuous features in $m(B)$. (c) 
Normalized magnetization measurements of Ref.~\cite{Kato_2017}. (d) Measured 
magnetization derivative. The features at the different fields $B_c$, where 
the gap is closed, and $B_{\rm sat}$, where saturation is achieved, demarcate 
two distinct magnetic phases. In panels (c) and (d) we show that quantitative 
agreement with experiment is obtained if the interaction parameters are scaled 
downwards by the factor $Z_c = 1.18$ arising from the corrections to LSW theory 
\cite{Igarashi92}.}
\label{fig:mag_curved}
\end{figure*}

\section{High-field magnetization}
\label{sec:mag}

The INS interaction parameters we obtain have direct implications for all 
of the magnetic, and by extension magnetoelectric, properties of \sample. 
The most detailed thermodynamic data available take the form of high-field 
magnetization measurements, which were performed in Ref.~\cite{Kato_2017} up 
to full saturation at fields in excess of 60 T. We repeat the CMF analysis 
of these data \cite{Kato_2017} using the parameters deduced from INS and the 
results are shown in Figs.~\ref{fig:mag_curved}(a) and \ref{fig:mag_curved}(b).

The magnetization measurements of Ref.~\cite{Kato_2017}, reproduced in 
Figs.~\ref{fig:mag_curved}(c) and \ref{fig:mag_curved}(d), are described to 
semi-quantitative accuracy by CMF modelling with the parameters of Table 
\ref{tab_Jparam}. This degree of consistency is eminently reassuring, and 
establishes the relative values of the INS parameters as the updated benchmark 
for \sample. However, given that the magnetization data were modelled 
previously with a much smaller $J_2'$, the question remains as to whether 
any given set of proposed parameters can be established uniquely. Here it is 
important to note that the two methods of analysis are quite different, as the 
CMF approach is based around the limit of weakly coupled plaquettes whereas 
LSW theory is based on the assumption of robust magnetic order throughout the 
system (meaning strongly coupled plaquettes). While a direct comparison is 
therefore not necessarily meaningful, our results provide an example in which 
their predictions agree rather well; LSW theory is justified by the strong 
$J_2'$ and the CMF method remains within its range of applicability because 
the net intra-plaquette interactions, $J_1 + D$ per bond, still exceed the 
inter-plaquette $J_2'$.

Qualitatively, this degree of consistency between the LSW and CMF descriptions 
of the magnetization suggests that there are no major discrepancies over issues 
such as the moment direction and the orientation or magnitude of the DM vector.
The CMF results show two magnetic phases below saturation, as deduced from the 
strongly anisotropic response to fields applied in different crystallographic 
directions (Fig.~\ref{fig:mag_curved}). In addition to these phases I and II, 
it was suggested \cite{Kato_2017} that the system may be close to a predicted 
phase III. However, no evidence for this possibility appears in purely magnetic 
measurements, and thus more sensitive dielectric measurements are required for 
a deeper analysis. One may also consider different materials in the family of 
$A$($B$O)Cu$_4$(PO$_4$)$_4$ compounds for a broader investigation of possible 
magnetic phases in this complex geometry. 

Quantitatively, the INS parameters appear to overestimate the experimental 
saturation fields (57--63 T for different field directions) by approximately 
20\% (Fig.~\ref{fig:mag_curved}). The saturation field is in general a 
coordination-weighted sum of all the $J_m$ parameters in Table \ref{tab_Jparam}.
We note that in a pure Heisenberg model on the 2D square lattice, the LSW 
approximation has been shown theoretically \cite{Igarashi92} to overestimate 
the values of $J_m$ by a factor $Z_c = 1.18$, as a consequence of the fact 
that it does not include quantum fluctuation corrections. The applicability 
of $Z_c$ has been verified in experiment \cite{Ronnow_2001}, and in 
Figs.~\ref{fig:mag_curved}(c) and \ref{fig:mag_curved}(d) we demonstrate that 
this correction alone is sufficient to achieve a quantitative match between 
our static and dynamic measurements.

\section{Discussion}
\label{sd}

The two key features of the parameter set we identify that defines the 
magnetic lattice in Ba(TiO)Cu$_4$(PO$_4$)$_4$ (Table \ref{tab_Jparam}) are 
the strength of the diagonal inter-plaquette coupling ($J_2' \approx J_1$) 
and the strong intra-plaquette DM interaction ($D \approx J_1/2$). As noted 
in Sec.~\ref{si}, the isotropic (Heisenberg) tetramerized $J_1$-$J_1'$ model 
has a quantum phase transition from a gapless magnetically ordered phase to 
a gapped plaquette-singlet phase at $\alpha_c = J_1'/J_1 \approx 0.55$ 
\cite{Albuquerque_2008,Wenzel2008}. At lowest order, the $J_1$-$J_2'$ model 
defined by the leading Heisenberg terms in Ba(TiO)Cu$_4$(PO$_4$)$_4$ would 
have the same behavior, with an inverted ordering pattern between plaquettes, 
in which case the $J_2'/J_1$ ratio would place the system well in the ordered 
phase. While further efforts have been applied to understanding the disordered 
phases arising in tetramerized square-lattice models \cite{Doretto2014,
Syromyatnikov2018,Syromyatnikov2020}, no anisotropy has yet been considered. 

In systems with DM interactions only on the inter-plaquette bonds, a 
gap is always present but a critical point remains, with $\alpha_c$ being 
determined from the onset of an ordered moment, and this type of physics 
has been discussed for the square-lattice dimer system Sr$_3$Ir$_2$O$_7$ 
\cite{Moretti_Sala_2015}. However, intra-plaquette DM interactions cause an 
admixture of triplets and quintuplets into the ground state at any finite 
$D$, as has been shown for coupled tetrahedra in the pyrochlore geometry 
\cite{Kotov2004}. The generic situation in a square-lattice model is then the 
immediate onset of long-range order in addition to the opening of a magnon gap, 
and as a result the $A$($B$O)Cu$_4$(PO$_4$)$_4$ family of compounds is pushed 
away from the phase space of gapped tetramerized $S = 1/2$ systems. 

For the geometry of \sample, the intra-plaquette spin configuration is that 
shown in Fig.~\ref{fig_magDM}(a), which is controlled by $D_\|$ rather than by 
$D_z$ (Table \ref{tab_Jparam}). The inter-plaquette alignment of cupola units 
presents no frustration problem when comparing $J_1'$ with $J_2'$ interactions, 
because the DM terms remain satisfied for either relative alignment of the 
$S_z$ spin components [Fig.~\ref{fig:structure}(c)]. Even in a system with 
very weak inter-plaquette coupling, the action of the intra-plaquette DM terms 
remains that of inducing a weak ordered moment, whose fluctuations are gapped 
magnon excitations, and these features are superposed upon the quantum 
fluctuation effects favoring isolated plaquette states. A heuristic measure 
of the influence of the strong DM interactions in \sample~on suppressing 
these quantum fluctuation effects can be obtained from the ordered moment, 
which was estimated from a detailed analysis of the magnetic structure at 
80\% of the maximal value \cite{Babkevich_2017}, as opposed to 61\% in an 
isotropic square lattice \cite{Manousakis_1991}.

Although the strong inter-plaquette interactions in \sample~place it rather
far from the parameter regime for investigating weak magnetic order coexisting 
with strong quantum fluctuations, they do make this material an excellent 
candidate for the study of topological magnon states \cite{Katsura10, 
Shindou13,Mook14}. It has been shown for both topological magnon 
\cite{vanHoogdalem13,Mook14} and triplon systems \cite{Romhanyi15} that  
the combination of multiple DM terms with an external magnetic field 
\cite{vanHoogdalem13,Romhanyi15} leads to symmetry-preserved topological 
modes that exhibit Dirac-cone level-crossing rather than level repulsion 
and anticrossing. We observe from Fig.~\ref{fig:IN12_field_cuts} that 
a field-driven gap closure, which is a candidate topological quantum 
phase transition, can be expected in \sample~at approximately 12 T 
(Fig.~\ref{fig:mag_curved}) \cite{Kato_2017}.  

\section{Summary}
\label{sec:conclusion}

We have measured the spin dynamics of the compound \sample, which is composed 
of Cu$_4$O$_{12}$ ``cupola'' units coupled into 2D square-lattice planes. Our 
high-quality data reveal a complex spectrum of well resolved magnetic 
excitations, whose evolution and systematic splitting we have followed to 
an applied magnetic field of 5~T. Despite the presence of robust magnetic 
order, the lowest-lying spin excitation has a large gap, of half its band 
width, indicating the presence of significant DM interactions. We obtain a 
quantitatively accurate description of every aspect of the measured spectrum 
by using a linear spin-wave theory, which indicates the primacy of the ordered 
moment in determining the appropriate description of the magnetic properties. 
Our fitted spectra indicate that the four-site plaquette units have strong 
intra-plaquette DM interactions ($D$) and strong inter-plaquette Heisenberg 
coupling ($J_2'$). The values we obtain, $D \simeq 0.53J_1$ and $J_2' \simeq 
1.09 J_1$, give results fully consistent with static measurements up to very 
high applied fields and thus set the benchmark for \sample. Although the 
minimal tetramerization and strong intra-plaquette DM interactions in the 
$A$($B$O)Cu$_4$(PO$_4$)$_4$ compounds suppress quantum fluctuation effects 
in favor of noncoplanar magnetic order, they also give this family of 
materials high potential for the systematic study of topological magnetic 
states and topological magnon excitations. 

\begin{acknowledgments}
We thank F. Groitl, F. Mila, and G. S. Tucker for helpful discussions. We are 
grateful to the Swiss National Science Foundation (SNSF) for financial support 
under Grant No.~188648, to the European Research Council (ERC) for the support 
of the Synergy network HERO (Grant No.~810451), to the Japan Society for the 
Promotion of Science (JSPS) for the support of KAKENHI Grants No.~JP19H05823 
and No.~JP19H01847, and for the support of the Japanese MEXT Leading Initiative 
for Excellent Young Researchers (LEADER). Access to MACS was provided by the 
Center for High Resolution Neutron Scattering, a partnership between the 
National Institute of Standards and Technology and the National Science 
Foundation under Agreement No.~DMR-1508249. Data collected at the Institut 
Laue-Langevin in the course of this study are available as 
Refs.~\cite{DOI_IN5,DOI_IN12}. The work performed on IN12 was supported 
by the Swiss State Secretariat for Education, Research and Innovation 
(SERI) through a CRG Grant. 
\end{acknowledgments}

\bibliographystyle{apsrev4-1}
\bibliography{cupolabib}

\begin{thebibliography}{62}%
\makeatletter
\providecommand \@ifxundefined [1]{%
 \@ifx{#1\undefined}
}%
\providecommand \@ifnum [1]{%
 \ifnum #1\expandafter \@firstoftwo
 \else \expandafter \@secondoftwo
 \fi
}%
\providecommand \@ifx [1]{%
 \ifx #1\expandafter \@firstoftwo
 \else \expandafter \@secondoftwo
 \fi
}%
\providecommand \natexlab [1]{#1}%
\providecommand \enquote  [1]{``#1''}%
\providecommand \bibnamefont  [1]{#1}%
\providecommand \bibfnamefont [1]{#1}%
\providecommand \citenamefont [1]{#1}%
\providecommand \href@noop [0]{\@secondoftwo}%
\providecommand \href [0]{\begingroup \@sanitize@url \@href}%
\providecommand \@href[1]{\@@startlink{#1}\@@href}%
\providecommand \@@href[1]{\endgroup#1\@@endlink}%
\providecommand \@sanitize@url [0]{\catcode `\\12\catcode `\$12\catcode
  `\&12\catcode `\#12\catcode `\^12\catcode `\_12\catcode `\%12\relax}%
\providecommand \@@startlink[1]{}%
\providecommand \@@endlink[0]{}%
\providecommand \url  [0]{\begingroup\@sanitize@url \@url }%
\providecommand \@url [1]{\endgroup\@href {#1}{\urlprefix }}%
\providecommand \urlprefix  [0]{URL }%
\providecommand \Eprint [0]{\href }%
\providecommand \doibase [0]{http://dx.doi.org/}%
\providecommand \selectlanguage [0]{\@gobble}%
\providecommand \bibinfo  [0]{\@secondoftwo}%
\providecommand \bibfield  [0]{\@secondoftwo}%
\providecommand \translation [1]{[#1]}%
\providecommand \BibitemOpen [0]{}%
\providecommand \bibitemStop [0]{}%
\providecommand \bibitemNoStop [0]{.\EOS\space}%
\providecommand \EOS [0]{\spacefactor3000\relax}%
\providecommand \BibitemShut  [1]{\csname bibitem#1\endcsname}%
\let\auto@bib@innerbib\@empty
\bibitem [{\citenamefont {Bethe}(1931)}]{Bethe_1931}%
  \BibitemOpen
  \bibfield  {author} {\bibinfo {author} {\bibfnamefont {H.}~\bibnamefont
  {Bethe}},\ }\href {\doibase 10.1007/BF01341708} {\bibfield  {journal}
  {\bibinfo  {journal} {Z. Phys.}\ }\textbf {\bibinfo {volume} {71}},\ \bibinfo
  {pages} {205} (\bibinfo {year} {1931})}\BibitemShut {NoStop}%
\bibitem [{\citenamefont {Faddeev}\ and\ \citenamefont
  {Takhtajan}(1981)}]{Faddeev_1981}%
  \BibitemOpen
  \bibfield  {author} {\bibinfo {author} {\bibfnamefont {L.~D.}\ \bibnamefont
  {Faddeev}}\ and\ \bibinfo {author} {\bibfnamefont {L.~A.}\ \bibnamefont
  {Takhtajan}},\ }\href {\doibase https://doi.org/10.1016/0375-9601(81)90335-2}
  {\bibfield  {journal} {\bibinfo  {journal} {Phys. Lett. A}\ }\textbf
  {\bibinfo {volume} {85}},\ \bibinfo {pages} {375 } (\bibinfo {year}
  {1981})}\BibitemShut {NoStop}%
\bibitem [{\citenamefont {M\"uller}\ \emph {et~al.}(1981)\citenamefont
  {M\"uller}, \citenamefont {Thomas}, \citenamefont {Beck},\ and\ \citenamefont
  {Bonner}}]{Muller_1981}%
  \BibitemOpen
  \bibfield  {author} {\bibinfo {author} {\bibfnamefont {G.}~\bibnamefont
  {M\"uller}}, \bibinfo {author} {\bibfnamefont {H.}~\bibnamefont {Thomas}},
  \bibinfo {author} {\bibfnamefont {H.}~\bibnamefont {Beck}}, \ and\ \bibinfo
  {author} {\bibfnamefont {J.~C.}\ \bibnamefont {Bonner}},\ }\href {\doibase
  10.1103/PhysRevB.24.1429} {\bibfield  {journal} {\bibinfo  {journal} {Phys.
  Rev. B}\ }\textbf {\bibinfo {volume} {24}},\ \bibinfo {pages} {1429}
  (\bibinfo {year} {1981})}\BibitemShut {NoStop}%
\bibitem [{\citenamefont {N\'eel}(1948)}]{Neel_1948}%
  \BibitemOpen
  \bibfield  {author} {\bibinfo {author} {\bibfnamefont {L.}~\bibnamefont
  {N\'eel}},\ }\href {\doibase 10.1051/anphys/194812030137} {\bibfield
  {journal} {\bibinfo  {journal} {Ann. Phys.}\ }\textbf {\bibinfo {volume}
  {12}},\ \bibinfo {pages} {137} (\bibinfo {year} {1948})}\BibitemShut
  {NoStop}%
\bibitem [{\citenamefont {Shull}\ \emph {et~al.}(1951)\citenamefont {Shull},
  \citenamefont {Strauser},\ and\ \citenamefont {Wollan}}]{Shull_1951}%
  \BibitemOpen
  \bibfield  {author} {\bibinfo {author} {\bibfnamefont {C.~G.}\ \bibnamefont
  {Shull}}, \bibinfo {author} {\bibfnamefont {W.~A.}\ \bibnamefont {Strauser}},
  \ and\ \bibinfo {author} {\bibfnamefont {E.~O.}\ \bibnamefont {Wollan}},\
  }\href {\doibase 10.1103/PhysRev.83.333} {\bibfield  {journal} {\bibinfo
  {journal} {Phys. Rev.}\ }\textbf {\bibinfo {volume} {83}},\ \bibinfo {pages}
  {333} (\bibinfo {year} {1951})}\BibitemShut {NoStop}%
\bibitem [{\citenamefont {Manousakis}(1991)}]{Manousakis_1991}%
  \BibitemOpen
  \bibfield  {author} {\bibinfo {author} {\bibfnamefont {E.}~\bibnamefont
  {Manousakis}},\ }\href {\doibase 10.1103/RevModPhys.63.1} {\bibfield
  {journal} {\bibinfo  {journal} {Rev. Mod. Phys.}\ }\textbf {\bibinfo {volume}
  {63}},\ \bibinfo {pages} {1} (\bibinfo {year} {1991})}\BibitemShut {NoStop}%
\bibitem [{\citenamefont {Anderson}(1973)}]{Anderson_1973}%
  \BibitemOpen
  \bibfield  {author} {\bibinfo {author} {\bibfnamefont {P.}~\bibnamefont
  {Anderson}},\ }\href {\doibase https://doi.org/10.1016/0025-5408(73)90167-0}
  {\bibfield  {journal} {\bibinfo  {journal} {Mat. Res. Bull.}\ }\textbf
  {\bibinfo {volume} {8}},\ \bibinfo {pages} {153 } (\bibinfo {year}
  {1973})}\BibitemShut {NoStop}%
\bibitem [{\citenamefont {Lee}\ \emph {et~al.}(2006)\citenamefont {Lee},
  \citenamefont {Nagaosa},\ and\ \citenamefont {Wen}}]{Lee2006}%
  \BibitemOpen
  \bibfield  {author} {\bibinfo {author} {\bibfnamefont {P.~A.}\ \bibnamefont
  {Lee}}, \bibinfo {author} {\bibfnamefont {N.}~\bibnamefont {Nagaosa}}, \ and\
  \bibinfo {author} {\bibfnamefont {X.-G.}\ \bibnamefont {Wen}},\ }\href
  {\doibase 10.1103/RevModPhys.78.17} {\bibfield  {journal} {\bibinfo
  {journal} {Rev. Mod. Phys.}\ }\textbf {\bibinfo {volume} {78}},\ \bibinfo
  {pages} {17} (\bibinfo {year} {2006})}\BibitemShut {NoStop}%
\bibitem [{\citenamefont {Chandra}\ and\ \citenamefont
  {Doucot}(1988)}]{Chandra_1988}%
  \BibitemOpen
  \bibfield  {author} {\bibinfo {author} {\bibfnamefont {P.}~\bibnamefont
  {Chandra}}\ and\ \bibinfo {author} {\bibfnamefont {B.}~\bibnamefont
  {Doucot}},\ }\href {\doibase 10.1103/PhysRevB.38.9335} {\bibfield  {journal}
  {\bibinfo  {journal} {Phys. Rev. B}\ }\textbf {\bibinfo {volume} {38}},\
  \bibinfo {pages} {9335} (\bibinfo {year} {1988})}\BibitemShut {NoStop}%
\bibitem [{\citenamefont {Balents}(2010)}]{Balents_2010}%
  \BibitemOpen
  \bibfield  {author} {\bibinfo {author} {\bibfnamefont {L.}~\bibnamefont
  {Balents}},\ }\href {\doibase 10.1038/nature08917} {\bibfield  {journal}
  {\bibinfo  {journal} {Nature}\ }\textbf {\bibinfo {volume} {464}},\ \bibinfo
  {pages} {199} (\bibinfo {year} {2010})}\BibitemShut {NoStop}%
\bibitem [{\citenamefont {Isaev}\ \emph {et~al.}(2009)\citenamefont {Isaev},
  \citenamefont {Ortiz},\ and\ \citenamefont
  {Dukelsky}}]{Isaev_2009_MF_square}%
  \BibitemOpen
  \bibfield  {author} {\bibinfo {author} {\bibfnamefont {L.}~\bibnamefont
  {Isaev}}, \bibinfo {author} {\bibfnamefont {G.}~\bibnamefont {Ortiz}}, \ and\
  \bibinfo {author} {\bibfnamefont {J.}~\bibnamefont {Dukelsky}},\ }\href
  {\doibase 10.1103/PhysRevB.79.024409} {\bibfield  {journal} {\bibinfo
  {journal} {Phys. Rev. B}\ }\textbf {\bibinfo {volume} {79}},\ \bibinfo
  {pages} {024409} (\bibinfo {year} {2009})}\BibitemShut {NoStop}%
\bibitem [{\citenamefont {G\"otze}\ \emph {et~al.}(2012)\citenamefont
  {G\"otze}, \citenamefont {Kr\"uger}, \citenamefont {Fleck}, \citenamefont
  {Schulenburg},\ and\ \citenamefont {Richter}}]{Gotze_2012}%
  \BibitemOpen
  \bibfield  {author} {\bibinfo {author} {\bibfnamefont {O.}~\bibnamefont
  {G\"otze}}, \bibinfo {author} {\bibfnamefont {S.~E.}\ \bibnamefont
  {Kr\"uger}}, \bibinfo {author} {\bibfnamefont {F.}~\bibnamefont {Fleck}},
  \bibinfo {author} {\bibfnamefont {J.}~\bibnamefont {Schulenburg}}, \ and\
  \bibinfo {author} {\bibfnamefont {J.}~\bibnamefont {Richter}},\ }\href
  {\doibase 10.1103/PhysRevB.85.224424} {\bibfield  {journal} {\bibinfo
  {journal} {Phys. Rev. B}\ }\textbf {\bibinfo {volume} {85}},\ \bibinfo
  {pages} {224424} (\bibinfo {year} {2012})}\BibitemShut {NoStop}%
\bibitem [{\citenamefont {Doretto}(2014)}]{Doretto2014}%
  \BibitemOpen
  \bibfield  {author} {\bibinfo {author} {\bibfnamefont {R.~L.}\ \bibnamefont
  {Doretto}},\ }\href {\doibase 10.1103/PhysRevB.89.104415} {\bibfield
  {journal} {\bibinfo  {journal} {Phys. Rev. B}\ }\textbf {\bibinfo {volume}
  {89}},\ \bibinfo {pages} {104415} (\bibinfo {year} {2014})}\BibitemShut
  {NoStop}%
\bibitem [{\citenamefont {Gong}\ \emph {et~al.}(2014)\citenamefont {Gong},
  \citenamefont {Zhu}, \citenamefont {Sheng}, \citenamefont {Motrunich},\ and\
  \citenamefont {Fisher}}]{Gong2014}%
  \BibitemOpen
  \bibfield  {author} {\bibinfo {author} {\bibfnamefont {S.-S.}\ \bibnamefont
  {Gong}}, \bibinfo {author} {\bibfnamefont {W.}~\bibnamefont {Zhu}}, \bibinfo
  {author} {\bibfnamefont {D.~N.}\ \bibnamefont {Sheng}}, \bibinfo {author}
  {\bibfnamefont {O.~I.}\ \bibnamefont {Motrunich}}, \ and\ \bibinfo {author}
  {\bibfnamefont {M.~P.~A.}\ \bibnamefont {Fisher}},\ }\href {\doibase
  10.1103/PhysRevLett.113.027201} {\bibfield  {journal} {\bibinfo  {journal}
  {Phys. Rev. Lett.}\ }\textbf {\bibinfo {volume} {113}},\ \bibinfo {pages}
  {027201} (\bibinfo {year} {2014})}\BibitemShut {NoStop}%
\bibitem [{\citenamefont {Morita}\ \emph {et~al.}(2015)\citenamefont {Morita},
  \citenamefont {Kaneko},\ and\ \citenamefont {Imada}}]{Morita2015}%
  \BibitemOpen
  \bibfield  {author} {\bibinfo {author} {\bibfnamefont {S.}~\bibnamefont
  {Morita}}, \bibinfo {author} {\bibfnamefont {R.}~\bibnamefont {Kaneko}}, \
  and\ \bibinfo {author} {\bibfnamefont {M.}~\bibnamefont {Imada}},\ }\href
  {\doibase 10.7566/JPSJ.84.024720} {\bibfield  {journal} {\bibinfo  {journal}
  {J. Phys. Soc. Jpn.}\ }\textbf {\bibinfo {volume} {84}},\ \bibinfo {pages}
  {024720} (\bibinfo {year} {2015})}\BibitemShut {NoStop}%
\bibitem [{\citenamefont {Wang}\ and\ \citenamefont
  {Sandvik}(2018)}]{Wang2018}%
  \BibitemOpen
  \bibfield  {author} {\bibinfo {author} {\bibfnamefont {L.}~\bibnamefont
  {Wang}}\ and\ \bibinfo {author} {\bibfnamefont {A.~W.}\ \bibnamefont
  {Sandvik}},\ }\href {\doibase 10.1103/PhysRevLett.121.107202} {\bibfield
  {journal} {\bibinfo  {journal} {Phys. Rev. Lett.}\ }\textbf {\bibinfo
  {volume} {121}},\ \bibinfo {pages} {107202} (\bibinfo {year}
  {2018})}\BibitemShut {NoStop}%
\bibitem [{\citenamefont {Haghshenas}\ and\ \citenamefont
  {Sheng}(2018)}]{Haghshenas2018}%
  \BibitemOpen
  \bibfield  {author} {\bibinfo {author} {\bibfnamefont {R.}~\bibnamefont
  {Haghshenas}}\ and\ \bibinfo {author} {\bibfnamefont {D.~N.}\ \bibnamefont
  {Sheng}},\ }\href {\doibase 10.1103/PhysRevB.97.1744408} {\bibfield
  {journal} {\bibinfo  {journal} {Phys. Rev. B}\ }\textbf {\bibinfo {volume}
  {97}},\ \bibinfo {pages} {174408} (\bibinfo {year} {2018})}\BibitemShut
  {NoStop}%
\bibitem [{\citenamefont {Albuquerque}\ \emph {et~al.}(2008)\citenamefont
  {Albuquerque}, \citenamefont {Troyer},\ and\ \citenamefont
  {Oitmaa}}]{Albuquerque_2008}%
  \BibitemOpen
  \bibfield  {author} {\bibinfo {author} {\bibfnamefont {A.~F.}\ \bibnamefont
  {Albuquerque}}, \bibinfo {author} {\bibfnamefont {M.}~\bibnamefont {Troyer}},
  \ and\ \bibinfo {author} {\bibfnamefont {J.}~\bibnamefont {Oitmaa}},\ }\href
  {\doibase 10.1103/PhysRevB.78.132402} {\bibfield  {journal} {\bibinfo
  {journal} {Phys. Rev. B}\ }\textbf {\bibinfo {volume} {78}},\ \bibinfo
  {pages} {132402} (\bibinfo {year} {2008})}\BibitemShut {NoStop}%
\bibitem [{\citenamefont {Wenzel}\ \emph {et~al.}(2008)\citenamefont {Wenzel},
  \citenamefont {Bogacz},\ and\ \citenamefont {Janke}}]{Wenzel2008}%
  \BibitemOpen
  \bibfield  {author} {\bibinfo {author} {\bibfnamefont {S.}~\bibnamefont
  {Wenzel}}, \bibinfo {author} {\bibfnamefont {L.}~\bibnamefont {Bogacz}}, \
  and\ \bibinfo {author} {\bibfnamefont {W.}~\bibnamefont {Janke}},\ }\href
  {\doibase 10.1103/PhysRevLett.101.127202} {\bibfield  {journal} {\bibinfo
  {journal} {Phys. Rev. Lett.}\ }\textbf {\bibinfo {volume} {101}},\ \bibinfo
  {pages} {127202} (\bibinfo {year} {2008})}\BibitemShut {NoStop}%
\bibitem [{\citenamefont {Bishop}\ \emph {et~al.}(2012)\citenamefont {Bishop},
  \citenamefont {Li}, \citenamefont {Farnell}, \citenamefont {Richter},\ and\
  \citenamefont {Campbell}}]{Bishop_2012}%
  \BibitemOpen
  \bibfield  {author} {\bibinfo {author} {\bibfnamefont {R.~F.}\ \bibnamefont
  {Bishop}}, \bibinfo {author} {\bibfnamefont {P.~H.~Y.}\ \bibnamefont {Li}},
  \bibinfo {author} {\bibfnamefont {D.~J.~J.}\ \bibnamefont {Farnell}},
  \bibinfo {author} {\bibfnamefont {J.}~\bibnamefont {Richter}}, \ and\
  \bibinfo {author} {\bibfnamefont {C.~E.}\ \bibnamefont {Campbell}},\ }\href
  {\doibase 10.1103/PhysRevB.85.205122} {\bibfield  {journal} {\bibinfo
  {journal} {Phys. Rev. B}\ }\textbf {\bibinfo {volume} {85}},\ \bibinfo
  {pages} {205122} (\bibinfo {year} {2012})}\BibitemShut {NoStop}%
\bibitem [{\citenamefont {Xu}\ \emph {et~al.}(2019)\citenamefont {Xu},
  \citenamefont {Xiong}, \citenamefont {Wu},\ and\ \citenamefont
  {Yao}}]{Xu_2019_QMC}%
  \BibitemOpen
  \bibfield  {author} {\bibinfo {author} {\bibfnamefont {Y.}~\bibnamefont
  {Xu}}, \bibinfo {author} {\bibfnamefont {Z.}~\bibnamefont {Xiong}}, \bibinfo
  {author} {\bibfnamefont {H.-Q.}\ \bibnamefont {Wu}}, \ and\ \bibinfo {author}
  {\bibfnamefont {D.-X.}\ \bibnamefont {Yao}},\ }\href {\doibase
  10.1103/PhysRevB.99.085112} {\bibfield  {journal} {\bibinfo  {journal} {Phys.
  Rev. B}\ }\textbf {\bibinfo {volume} {99}},\ \bibinfo {pages} {085112}
  (\bibinfo {year} {2019})}\BibitemShut {NoStop}%
\bibitem [{\citenamefont {Dzyaloshinsky}(1958)}]{Dzyaloshinskii_1958}%
  \BibitemOpen
  \bibfield  {author} {\bibinfo {author} {\bibfnamefont {I.}~\bibnamefont
  {Dzyaloshinsky}},\ }\href {\doibase
  https://doi.org/10.1016/0022-3697(58)90076-3} {\bibfield  {journal} {\bibinfo
   {journal} {J. Phys. Chem. Sol.}\ }\textbf {\bibinfo {volume} {4}},\ \bibinfo
  {pages} {241} (\bibinfo {year} {1958})}\BibitemShut {NoStop}%
\bibitem [{\citenamefont {Moriya}(1960)}]{Moriya_1960}%
  \BibitemOpen
  \bibfield  {author} {\bibinfo {author} {\bibfnamefont {T.}~\bibnamefont
  {Moriya}},\ }\href {\doibase 10.1103/PhysRev.120.91} {\bibfield  {journal}
  {\bibinfo  {journal} {Phys. Rev.}\ }\textbf {\bibinfo {volume} {120}},\
  \bibinfo {pages} {91} (\bibinfo {year} {1960})}\BibitemShut {NoStop}%
\bibitem [{\citenamefont {Miyahara}\ \emph {et~al.}(2007)\citenamefont
  {Miyahara}, \citenamefont {Fouet}, \citenamefont {Manmana}, \citenamefont
  {Noack}, \citenamefont {Mayaffre}, \citenamefont {Sheikin}, \citenamefont
  {Berthier},\ and\ \citenamefont {Mila}}]{Miyahara2007}%
  \BibitemOpen
  \bibfield  {author} {\bibinfo {author} {\bibfnamefont {S.}~\bibnamefont
  {Miyahara}}, \bibinfo {author} {\bibfnamefont {J.-B.}\ \bibnamefont {Fouet}},
  \bibinfo {author} {\bibfnamefont {S.~R.}\ \bibnamefont {Manmana}}, \bibinfo
  {author} {\bibfnamefont {R.~M.}\ \bibnamefont {Noack}}, \bibinfo {author}
  {\bibfnamefont {H.}~\bibnamefont {Mayaffre}}, \bibinfo {author}
  {\bibfnamefont {I.}~\bibnamefont {Sheikin}}, \bibinfo {author} {\bibfnamefont
  {C.}~\bibnamefont {Berthier}}, \ and\ \bibinfo {author} {\bibfnamefont
  {F.}~\bibnamefont {Mila}},\ }\href {\doibase 10.1103/PhysRevB.75.184402}
  {\bibfield  {journal} {\bibinfo  {journal} {Phys. Rev. B}\ }\textbf {\bibinfo
  {volume} {75}},\ \bibinfo {pages} {184402} (\bibinfo {year}
  {2007})}\BibitemShut {NoStop}%
\bibitem [{\citenamefont {Kotov}\ \emph {et~al.}(2004)\citenamefont {Kotov},
  \citenamefont {Zhitomirsky}, \citenamefont {Elhajal},\ and\ \citenamefont
  {Mila}}]{Kotov2004}%
  \BibitemOpen
  \bibfield  {author} {\bibinfo {author} {\bibfnamefont {V.~N.}\ \bibnamefont
  {Kotov}}, \bibinfo {author} {\bibfnamefont {M.~E.}\ \bibnamefont
  {Zhitomirsky}}, \bibinfo {author} {\bibfnamefont {M.}~\bibnamefont
  {Elhajal}}, \ and\ \bibinfo {author} {\bibfnamefont {F.}~\bibnamefont
  {Mila}},\ }\href {\doibase 10.1103/PhysRevB.70.214401} {\bibfield  {journal}
  {\bibinfo  {journal} {Phys. Rev. B}\ }\textbf {\bibinfo {volume} {70}},\
  \bibinfo {pages} {214401} (\bibinfo {year} {2004})}\BibitemShut {NoStop}%
\bibitem [{\citenamefont {Xi}\ \emph {et~al.}(2011)\citenamefont {Xi},
  \citenamefont {Hu}, \citenamefont {Zhao}, \citenamefont {Su}, \citenamefont
  {Normand},\ and\ \citenamefont {Wang}}]{Xi2011}%
  \BibitemOpen
  \bibfield  {author} {\bibinfo {author} {\bibfnamefont {B.}~\bibnamefont
  {Xi}}, \bibinfo {author} {\bibfnamefont {S.}~\bibnamefont {Hu}}, \bibinfo
  {author} {\bibfnamefont {J.}~\bibnamefont {Zhao}}, \bibinfo {author}
  {\bibfnamefont {G.}~\bibnamefont {Su}}, \bibinfo {author} {\bibfnamefont
  {B.}~\bibnamefont {Normand}}, \ and\ \bibinfo {author} {\bibfnamefont
  {X.}~\bibnamefont {Wang}},\ }\href {\doibase 10.1103/PhysRevB.84.134407}
  {\bibfield  {journal} {\bibinfo  {journal} {Phys. Rev. B}\ }\textbf {\bibinfo
  {volume} {84}},\ \bibinfo {pages} {134407} (\bibinfo {year}
  {2011})}\BibitemShut {NoStop}%
\bibitem [{\citenamefont {Rufo}\ \emph {et~al.}(2019)\citenamefont {Rufo},
  \citenamefont {de~Sousa},\ and\ \citenamefont {Plascak}}]{Rufo2019}%
  \BibitemOpen
  \bibfield  {author} {\bibinfo {author} {\bibfnamefont {S.}~\bibnamefont
  {Rufo}}, \bibinfo {author} {\bibfnamefont {J.~R.}\ \bibnamefont {de~Sousa}},
  \ and\ \bibinfo {author} {\bibfnamefont {J.~A.}\ \bibnamefont {Plascak}},\
  }\href {\doibase https://doi.org/10.1016/j.physa.2018.12.015} {\bibfield
  {journal} {\bibinfo  {journal} {Physica A}\ }\textbf {\bibinfo {volume}
  {518}},\ \bibinfo {pages} {349 } (\bibinfo {year} {2019})}\BibitemShut
  {NoStop}%
\bibitem [{\citenamefont {Coldea}\ \emph {et~al.}(2001)\citenamefont {Coldea},
  \citenamefont {Hayden}, \citenamefont {Aeppli}, \citenamefont {Perring},
  \citenamefont {Frost}, \citenamefont {Mason}, \citenamefont {Cheong},\ and\
  \citenamefont {Fisk}}]{Coldea2001}%
  \BibitemOpen
  \bibfield  {author} {\bibinfo {author} {\bibfnamefont {R.}~\bibnamefont
  {Coldea}}, \bibinfo {author} {\bibfnamefont {S.~M.}\ \bibnamefont {Hayden}},
  \bibinfo {author} {\bibfnamefont {G.}~\bibnamefont {Aeppli}}, \bibinfo
  {author} {\bibfnamefont {T.~G.}\ \bibnamefont {Perring}}, \bibinfo {author}
  {\bibfnamefont {C.~D.}\ \bibnamefont {Frost}}, \bibinfo {author}
  {\bibfnamefont {T.~E.}\ \bibnamefont {Mason}}, \bibinfo {author}
  {\bibfnamefont {S.-W.}\ \bibnamefont {Cheong}}, \ and\ \bibinfo {author}
  {\bibfnamefont {Z.}~\bibnamefont {Fisk}},\ }\href {\doibase
  10.1103/PhysRevLett.86.5377} {\bibfield  {journal} {\bibinfo  {journal}
  {Phys. Rev. Lett.}\ }\textbf {\bibinfo {volume} {86}},\ \bibinfo {pages}
  {5377} (\bibinfo {year} {2001})}\BibitemShut {NoStop}%
\bibitem [{\citenamefont {Headings}\ \emph {et~al.}(2010)\citenamefont
  {Headings}, \citenamefont {Hayden}, \citenamefont {Coldea},\ and\
  \citenamefont {Perring}}]{Headings2010}%
  \BibitemOpen
  \bibfield  {author} {\bibinfo {author} {\bibfnamefont {N.~S.}\ \bibnamefont
  {Headings}}, \bibinfo {author} {\bibfnamefont {S.~M.}\ \bibnamefont
  {Hayden}}, \bibinfo {author} {\bibfnamefont {R.}~\bibnamefont {Coldea}}, \
  and\ \bibinfo {author} {\bibfnamefont {T.~G.}\ \bibnamefont {Perring}},\
  }\href {\doibase 10.1103/PhysRevLett.105.247001} {\bibfield  {journal}
  {\bibinfo  {journal} {Phys. Rev. Lett.}\ }\textbf {\bibinfo {volume} {105}},\
  \bibinfo {pages} {247001} (\bibinfo {year} {2010})}\BibitemShut {NoStop}%
\bibitem [{\citenamefont {R\o{}nnow}\ \emph {et~al.}(2001)\citenamefont
  {R\o{}nnow}, \citenamefont {McMorrow}, \citenamefont {Coldea}, \citenamefont
  {Harrison}, \citenamefont {Youngson}, \citenamefont {Perring}, \citenamefont
  {Aeppli}, \citenamefont {Sylju\aa{}sen}, \citenamefont {Lefmann},\ and\
  \citenamefont {Rischel}}]{Ronnow_2001}%
  \BibitemOpen
  \bibfield  {author} {\bibinfo {author} {\bibfnamefont {H.~M.}\ \bibnamefont
  {R\o{}nnow}}, \bibinfo {author} {\bibfnamefont {D.~F.}\ \bibnamefont
  {McMorrow}}, \bibinfo {author} {\bibfnamefont {R.}~\bibnamefont {Coldea}},
  \bibinfo {author} {\bibfnamefont {A.}~\bibnamefont {Harrison}}, \bibinfo
  {author} {\bibfnamefont {I.~D.}\ \bibnamefont {Youngson}}, \bibinfo {author}
  {\bibfnamefont {T.~G.}\ \bibnamefont {Perring}}, \bibinfo {author}
  {\bibfnamefont {G.}~\bibnamefont {Aeppli}}, \bibinfo {author} {\bibfnamefont
  {O.}~\bibnamefont {Sylju\aa{}sen}}, \bibinfo {author} {\bibfnamefont
  {K.}~\bibnamefont {Lefmann}}, \ and\ \bibinfo {author} {\bibfnamefont
  {C.}~\bibnamefont {Rischel}},\ }\href {\doibase
  10.1103/PhysRevLett.87.037202} {\bibfield  {journal} {\bibinfo  {journal}
  {Phys. Rev. Lett.}\ }\textbf {\bibinfo {volume} {87}},\ \bibinfo {pages}
  {037202} (\bibinfo {year} {2001})}\BibitemShut {NoStop}%
\bibitem [{\citenamefont {Babkevich}\ \emph
  {et~al.}(2016{\natexlab{a}})\citenamefont {Babkevich}, \citenamefont
  {Katukuri}, \citenamefont {F\aa{}k}, \citenamefont {Rols}, \citenamefont
  {Fennell}, \citenamefont {Paji\'{c}}, \citenamefont {Tanaka}, \citenamefont
  {Pardini}, \citenamefont {Singh}, \citenamefont {Mitrushchenkov},
  \citenamefont {Yazyev},\ and\ \citenamefont {R\o{}nnow}}]{Babkevich1016}%
  \BibitemOpen
  \bibfield  {author} {\bibinfo {author} {\bibfnamefont {P.}~\bibnamefont
  {Babkevich}}, \bibinfo {author} {\bibfnamefont {V.~M.}\ \bibnamefont
  {Katukuri}}, \bibinfo {author} {\bibfnamefont {B.}~\bibnamefont {F\aa{}k}},
  \bibinfo {author} {\bibfnamefont {S.}~\bibnamefont {Rols}}, \bibinfo {author}
  {\bibfnamefont {T.}~\bibnamefont {Fennell}}, \bibinfo {author} {\bibfnamefont
  {D.}~\bibnamefont {Paji\'{c}}}, \bibinfo {author} {\bibfnamefont
  {H.}~\bibnamefont {Tanaka}}, \bibinfo {author} {\bibfnamefont
  {T.}~\bibnamefont {Pardini}}, \bibinfo {author} {\bibfnamefont {R.~R.~P.}\
  \bibnamefont {Singh}}, \bibinfo {author} {\bibfnamefont {A.}~\bibnamefont
  {Mitrushchenkov}}, \bibinfo {author} {\bibfnamefont {O.~V.}\ \bibnamefont
  {Yazyev}}, \ and\ \bibinfo {author} {\bibfnamefont {H.~M.}\ \bibnamefont
  {R\o{}nnow}},\ }\href {\doibase 10.1103/PhysRevLett.117.237203} {\bibfield
  {journal} {\bibinfo  {journal} {Phys. Rev. Lett.}\ }\textbf {\bibinfo
  {volume} {117}},\ \bibinfo {pages} {237203} (\bibinfo {year}
  {2016}{\natexlab{a}})}\BibitemShut {NoStop}%
\bibitem [{\citenamefont {Tsirlin}\ and\ \citenamefont
  {Rosner}(2009)}]{Tsirlin_2009}%
  \BibitemOpen
  \bibfield  {author} {\bibinfo {author} {\bibfnamefont {A.~A.}\ \bibnamefont
  {Tsirlin}}\ and\ \bibinfo {author} {\bibfnamefont {H.}~\bibnamefont
  {Rosner}},\ }\href {\doibase 10.1103/PhysRevB.79.214417} {\bibfield
  {journal} {\bibinfo  {journal} {Phys. Rev. B}\ }\textbf {\bibinfo {volume}
  {79}},\ \bibinfo {pages} {214417} (\bibinfo {year} {2009})}\BibitemShut
  {NoStop}%
\bibitem [{\citenamefont {Tsirlin}\ \emph {et~al.}(2011)\citenamefont
  {Tsirlin}, \citenamefont {Nath}, \citenamefont {Abakumov}, \citenamefont
  {Furukawa}, \citenamefont {Johnston}, \citenamefont {Hemmida}, \citenamefont
  {Krug~von Nidda}, \citenamefont {Loidl}, \citenamefont {Geibel},\ and\
  \citenamefont {Rosner}}]{Tsirlin_2011}%
  \BibitemOpen
  \bibfield  {author} {\bibinfo {author} {\bibfnamefont {A.~A.}\ \bibnamefont
  {Tsirlin}}, \bibinfo {author} {\bibfnamefont {R.}~\bibnamefont {Nath}},
  \bibinfo {author} {\bibfnamefont {A.~M.}\ \bibnamefont {Abakumov}}, \bibinfo
  {author} {\bibfnamefont {Y.}~\bibnamefont {Furukawa}}, \bibinfo {author}
  {\bibfnamefont {D.~C.}\ \bibnamefont {Johnston}}, \bibinfo {author}
  {\bibfnamefont {M.}~\bibnamefont {Hemmida}}, \bibinfo {author} {\bibfnamefont
  {H.-A.}\ \bibnamefont {Krug~von Nidda}}, \bibinfo {author} {\bibfnamefont
  {A.}~\bibnamefont {Loidl}}, \bibinfo {author} {\bibfnamefont
  {C.}~\bibnamefont {Geibel}}, \ and\ \bibinfo {author} {\bibfnamefont
  {H.}~\bibnamefont {Rosner}},\ }\href {\doibase 10.1103/PhysRevB.84.014429}
  {\bibfield  {journal} {\bibinfo  {journal} {Phys. Rev. B}\ }\textbf {\bibinfo
  {volume} {84}},\ \bibinfo {pages} {014429} (\bibinfo {year}
  {2011})}\BibitemShut {NoStop}%
\bibitem [{\citenamefont {Zhu}\ \emph {et~al.}(2010)\citenamefont {Zhu},
  \citenamefont {Yu}, \citenamefont {Wang}, \citenamefont {Zhao}, \citenamefont
  {Jones}, \citenamefont {Dai}, \citenamefont {Abrahams}, \citenamefont
  {Morosan}, \citenamefont {Fang},\ and\ \citenamefont {Si}}]{Zhu_2010}%
  \BibitemOpen
  \bibfield  {author} {\bibinfo {author} {\bibfnamefont {J.-X.}\ \bibnamefont
  {Zhu}}, \bibinfo {author} {\bibfnamefont {R.}~\bibnamefont {Yu}}, \bibinfo
  {author} {\bibfnamefont {H.}~\bibnamefont {Wang}}, \bibinfo {author}
  {\bibfnamefont {L.~L.}\ \bibnamefont {Zhao}}, \bibinfo {author}
  {\bibfnamefont {M.~D.}\ \bibnamefont {Jones}}, \bibinfo {author}
  {\bibfnamefont {J.}~\bibnamefont {Dai}}, \bibinfo {author} {\bibfnamefont
  {E.}~\bibnamefont {Abrahams}}, \bibinfo {author} {\bibfnamefont
  {E.}~\bibnamefont {Morosan}}, \bibinfo {author} {\bibfnamefont
  {M.}~\bibnamefont {Fang}}, \ and\ \bibinfo {author} {\bibfnamefont
  {Q.}~\bibnamefont {Si}},\ }\href {\doibase 10.1103/PhysRevLett.104.216405}
  {\bibfield  {journal} {\bibinfo  {journal} {Phys. Rev. Lett.}\ }\textbf
  {\bibinfo {volume} {104}},\ \bibinfo {pages} {216405} (\bibinfo {year}
  {2010})}\BibitemShut {NoStop}%
\bibitem [{\citenamefont {Kimura}\ \emph
  {et~al.}(2016{\natexlab{a}})\citenamefont {Kimura}, \citenamefont {Sera},\
  and\ \citenamefont {Kimura}}]{Kimura_2016_inorgchem}%
  \BibitemOpen
  \bibfield  {author} {\bibinfo {author} {\bibfnamefont {K.}~\bibnamefont
  {Kimura}}, \bibinfo {author} {\bibfnamefont {M.}~\bibnamefont {Sera}}, \ and\
  \bibinfo {author} {\bibfnamefont {T.}~\bibnamefont {Kimura}},\ }\href
  {\doibase 10.1021/acs.inorgchem.5b02622} {\bibfield  {journal} {\bibinfo
  {journal} {Inorg. Chem.}\ }\textbf {\bibinfo {volume} {55}},\ \bibinfo
  {pages} {1002} (\bibinfo {year} {2016}{\natexlab{a}})}\BibitemShut {NoStop}%
\bibitem [{\citenamefont {Kimura}\ \emph
  {et~al.}(2020{\natexlab{a}})\citenamefont {Kimura}, \citenamefont
  {Urushihara}, \citenamefont {Asaka}, \citenamefont {Toyoda}, \citenamefont
  {Miyake}, \citenamefont {Tokunaga}, \citenamefont {Matsuo}, \citenamefont
  {Kindo}, \citenamefont {Yamauchi},\ and\ \citenamefont
  {Kimura}}]{Kimura_2020_acs}%
  \BibitemOpen
  \bibfield  {author} {\bibinfo {author} {\bibfnamefont {K.}~\bibnamefont
  {Kimura}}, \bibinfo {author} {\bibfnamefont {D.}~\bibnamefont {Urushihara}},
  \bibinfo {author} {\bibfnamefont {T.}~\bibnamefont {Asaka}}, \bibinfo
  {author} {\bibfnamefont {M.}~\bibnamefont {Toyoda}}, \bibinfo {author}
  {\bibfnamefont {A.}~\bibnamefont {Miyake}}, \bibinfo {author} {\bibfnamefont
  {M.}~\bibnamefont {Tokunaga}}, \bibinfo {author} {\bibfnamefont
  {A.}~\bibnamefont {Matsuo}}, \bibinfo {author} {\bibfnamefont
  {K.}~\bibnamefont {Kindo}}, \bibinfo {author} {\bibfnamefont
  {K.}~\bibnamefont {Yamauchi}}, \ and\ \bibinfo {author} {\bibfnamefont
  {T.}~\bibnamefont {Kimura}},\ }\href {\doibase 10.1021/acs.inorgchem.0c01463}
  {\bibfield  {journal} {\bibinfo  {journal} {Inorg. Chem.}\ }\textbf {\bibinfo
  {volume} {59}},\ \bibinfo {pages} {10986} (\bibinfo {year}
  {2020}{\natexlab{a}})}\BibitemShut {NoStop}%
\bibitem [{\citenamefont {Kimura}\ \emph
  {et~al.}(2016{\natexlab{b}})\citenamefont {Kimura}, \citenamefont
  {Babkevich}, \citenamefont {Sera}, \citenamefont {Toyoda}, \citenamefont
  {Yamauchi}, \citenamefont {Tucker}, \citenamefont {Martius}, \citenamefont
  {Fennell}, \citenamefont {Manuel}, \citenamefont {Khalyavin}, \citenamefont
  {Johnson}, \citenamefont {Nakano}, \citenamefont {Nozue}, \citenamefont
  {R{\o}nnow},\ and\ \citenamefont {Kimura}}]{Kimura_Ncomms_2016}%
  \BibitemOpen
  \bibfield  {author} {\bibinfo {author} {\bibfnamefont {K.}~\bibnamefont
  {Kimura}}, \bibinfo {author} {\bibfnamefont {P.}~\bibnamefont {Babkevich}},
  \bibinfo {author} {\bibfnamefont {M.}~\bibnamefont {Sera}}, \bibinfo {author}
  {\bibfnamefont {M.}~\bibnamefont {Toyoda}}, \bibinfo {author} {\bibfnamefont
  {K.}~\bibnamefont {Yamauchi}}, \bibinfo {author} {\bibfnamefont {G.~S.}\
  \bibnamefont {Tucker}}, \bibinfo {author} {\bibfnamefont {J.}~\bibnamefont
  {Martius}}, \bibinfo {author} {\bibfnamefont {T.}~\bibnamefont {Fennell}},
  \bibinfo {author} {\bibfnamefont {P.}~\bibnamefont {Manuel}}, \bibinfo
  {author} {\bibfnamefont {D.~D.}\ \bibnamefont {Khalyavin}}, \bibinfo {author}
  {\bibfnamefont {R.~D.}\ \bibnamefont {Johnson}}, \bibinfo {author}
  {\bibfnamefont {T.}~\bibnamefont {Nakano}}, \bibinfo {author} {\bibfnamefont
  {Y.}~\bibnamefont {Nozue}}, \bibinfo {author} {\bibfnamefont {H.~M.}\
  \bibnamefont {R{\o}nnow}}, \ and\ \bibinfo {author} {\bibfnamefont
  {T.}~\bibnamefont {Kimura}},\ }\href {\doibase 10.1038/ncomms13039}
  {\bibfield  {journal} {\bibinfo  {journal} {Nat. Commun.}\ }\textbf {\bibinfo
  {volume} {7}},\ \bibinfo {pages} {13039} (\bibinfo {year}
  {2016}{\natexlab{b}})}\BibitemShut {NoStop}%
\bibitem [{\citenamefont {Kato}\ \emph {et~al.}(2017)\citenamefont {Kato},
  \citenamefont {Kimura}, \citenamefont {Miyake}, \citenamefont {Tokunaga},
  \citenamefont {Matsuo}, \citenamefont {Kindo}, \citenamefont {Akaki},
  \citenamefont {Hagiwara}, \citenamefont {Sera}, \citenamefont {Kimura},\ and\
  \citenamefont {Motome}}]{Kato_2017}%
  \BibitemOpen
  \bibfield  {author} {\bibinfo {author} {\bibfnamefont {Y.}~\bibnamefont
  {Kato}}, \bibinfo {author} {\bibfnamefont {K.}~\bibnamefont {Kimura}},
  \bibinfo {author} {\bibfnamefont {A.}~\bibnamefont {Miyake}}, \bibinfo
  {author} {\bibfnamefont {M.}~\bibnamefont {Tokunaga}}, \bibinfo {author}
  {\bibfnamefont {A.}~\bibnamefont {Matsuo}}, \bibinfo {author} {\bibfnamefont
  {K.}~\bibnamefont {Kindo}}, \bibinfo {author} {\bibfnamefont
  {M.}~\bibnamefont {Akaki}}, \bibinfo {author} {\bibfnamefont
  {M.}~\bibnamefont {Hagiwara}}, \bibinfo {author} {\bibfnamefont
  {M.}~\bibnamefont {Sera}}, \bibinfo {author} {\bibfnamefont {T.}~\bibnamefont
  {Kimura}}, \ and\ \bibinfo {author} {\bibfnamefont {Y.}~\bibnamefont
  {Motome}},\ }\href {\doibase 10.1103/PhysRevLett.118.107601} {\bibfield
  {journal} {\bibinfo  {journal} {Phys. Rev. Lett.}\ }\textbf {\bibinfo
  {volume} {118}},\ \bibinfo {pages} {107601} (\bibinfo {year}
  {2017})}\BibitemShut {NoStop}%
\bibitem [{\citenamefont {Kimura}\ \emph
  {et~al.}(2018{\natexlab{a}})\citenamefont {Kimura}, \citenamefont {Toyoda},
  \citenamefont {Babkevich}, \citenamefont {Yamauchi}, \citenamefont {Sera},
  \citenamefont {Nassif}, \citenamefont {R\o{}nnow},\ and\ \citenamefont
  {Kimura}}]{Kimura_2018}%
  \BibitemOpen
  \bibfield  {author} {\bibinfo {author} {\bibfnamefont {K.}~\bibnamefont
  {Kimura}}, \bibinfo {author} {\bibfnamefont {M.}~\bibnamefont {Toyoda}},
  \bibinfo {author} {\bibfnamefont {P.}~\bibnamefont {Babkevich}}, \bibinfo
  {author} {\bibfnamefont {K.}~\bibnamefont {Yamauchi}}, \bibinfo {author}
  {\bibfnamefont {M.}~\bibnamefont {Sera}}, \bibinfo {author} {\bibfnamefont
  {V.}~\bibnamefont {Nassif}}, \bibinfo {author} {\bibfnamefont {H.~M.}\
  \bibnamefont {R\o{}nnow}}, \ and\ \bibinfo {author} {\bibfnamefont
  {T.}~\bibnamefont {Kimura}},\ }\href {\doibase 10.1103/PhysRevB.97.134418}
  {\bibfield  {journal} {\bibinfo  {journal} {Phys. Rev. B}\ }\textbf {\bibinfo
  {volume} {97}},\ \bibinfo {pages} {134418} (\bibinfo {year}
  {2018}{\natexlab{a}})}\BibitemShut {NoStop}%
\bibitem [{\citenamefont {Islam}\ \emph {et~al.}(2018)\citenamefont {Islam},
  \citenamefont {Ranjith}, \citenamefont {Baenitz}, \citenamefont {Skourski},
  \citenamefont {Tsirlin},\ and\ \citenamefont {Nath}}]{Islam_2018}%
  \BibitemOpen
  \bibfield  {author} {\bibinfo {author} {\bibfnamefont {S.~S.}\ \bibnamefont
  {Islam}}, \bibinfo {author} {\bibfnamefont {K.~M.}\ \bibnamefont {Ranjith}},
  \bibinfo {author} {\bibfnamefont {M.}~\bibnamefont {Baenitz}}, \bibinfo
  {author} {\bibfnamefont {Y.}~\bibnamefont {Skourski}}, \bibinfo {author}
  {\bibfnamefont {A.~A.}\ \bibnamefont {Tsirlin}}, \ and\ \bibinfo {author}
  {\bibfnamefont {R.}~\bibnamefont {Nath}},\ }\href {\doibase
  https://doi.org/10.1103/PhysRevB.97.174432} {\bibfield  {journal} {\bibinfo
  {journal} {Phys. Rev. B}\ }\textbf {\bibinfo {volume} {97}},\ \bibinfo
  {pages} {174432} (\bibinfo {year} {2018})}\BibitemShut {NoStop}%
\bibitem [{\citenamefont {Kato}\ \emph {et~al.}(2019)\citenamefont {Kato},
  \citenamefont {Kimura}, \citenamefont {Miyake}, \citenamefont {Tokunaga},
  \citenamefont {Matsuo}, \citenamefont {Kindo}, \citenamefont {Akaki},
  \citenamefont {Hagiwara}, \citenamefont {Kimura}, \citenamefont {Kimura},\
  and\ \citenamefont {Motome}}]{Kato_2019}%
  \BibitemOpen
  \bibfield  {author} {\bibinfo {author} {\bibfnamefont {Y.}~\bibnamefont
  {Kato}}, \bibinfo {author} {\bibfnamefont {K.}~\bibnamefont {Kimura}},
  \bibinfo {author} {\bibfnamefont {A.}~\bibnamefont {Miyake}}, \bibinfo
  {author} {\bibfnamefont {M.}~\bibnamefont {Tokunaga}}, \bibinfo {author}
  {\bibfnamefont {A.}~\bibnamefont {Matsuo}}, \bibinfo {author} {\bibfnamefont
  {K.}~\bibnamefont {Kindo}}, \bibinfo {author} {\bibfnamefont
  {M.}~\bibnamefont {Akaki}}, \bibinfo {author} {\bibfnamefont
  {M.}~\bibnamefont {Hagiwara}}, \bibinfo {author} {\bibfnamefont
  {S.}~\bibnamefont {Kimura}}, \bibinfo {author} {\bibfnamefont
  {T.}~\bibnamefont {Kimura}}, \ and\ \bibinfo {author} {\bibfnamefont
  {Y.}~\bibnamefont {Motome}},\ }\href {\doibase 10.1103/PhysRevB.99.024415}
  {\bibfield  {journal} {\bibinfo  {journal} {Phys. Rev. B}\ }\textbf {\bibinfo
  {volume} {99}},\ \bibinfo {pages} {024415} (\bibinfo {year}
  {2019})}\BibitemShut {NoStop}%
\bibitem [{\citenamefont {Kumar}\ \emph {et~al.}(2019)\citenamefont {Kumar},
  \citenamefont {Shahee}, \citenamefont {Kundu}, \citenamefont {Baenitz},\ and\
  \citenamefont {Mahajan}}]{Kumar19}%
  \BibitemOpen
  \bibfield  {author} {\bibinfo {author} {\bibfnamefont {V.}~\bibnamefont
  {Kumar}}, \bibinfo {author} {\bibfnamefont {A.}~\bibnamefont {Shahee}},
  \bibinfo {author} {\bibfnamefont {S.}~\bibnamefont {Kundu}}, \bibinfo
  {author} {\bibfnamefont {M.}~\bibnamefont {Baenitz}}, \ and\ \bibinfo
  {author} {\bibfnamefont {A.~V.}\ \bibnamefont {Mahajan}},\ }\href {\doibase
  10.1016/j.jmmm.2019.165600} {\bibfield  {journal} {\bibinfo  {journal} {J.
  Magn. Magn. Mater.}\ }\textbf {\bibinfo {volume} {492}},\ \bibinfo {pages}
  {165600} (\bibinfo {year} {2019})}\BibitemShut {NoStop}%
\bibitem [{\citenamefont {Babkevich}\ \emph {et~al.}(2017)\citenamefont
  {Babkevich}, \citenamefont {Testa}, \citenamefont {Kimura}, \citenamefont
  {Kimura}, \citenamefont {Tucker}, \citenamefont {Roessli},\ and\
  \citenamefont {R\o{}nnow}}]{Babkevich_2017}%
  \BibitemOpen
  \bibfield  {author} {\bibinfo {author} {\bibfnamefont {P.}~\bibnamefont
  {Babkevich}}, \bibinfo {author} {\bibfnamefont {L.}~\bibnamefont {Testa}},
  \bibinfo {author} {\bibfnamefont {K.}~\bibnamefont {Kimura}}, \bibinfo
  {author} {\bibfnamefont {T.}~\bibnamefont {Kimura}}, \bibinfo {author}
  {\bibfnamefont {G.~S.}\ \bibnamefont {Tucker}}, \bibinfo {author}
  {\bibfnamefont {B.}~\bibnamefont {Roessli}}, \ and\ \bibinfo {author}
  {\bibfnamefont {H.~M.}\ \bibnamefont {R\o{}nnow}},\ }\href {\doibase
  10.1103/PhysRevB.96.214436} {\bibfield  {journal} {\bibinfo  {journal} {Phys.
  Rev. B}\ }\textbf {\bibinfo {volume} {96}},\ \bibinfo {pages} {214436}
  (\bibinfo {year} {2017})}\BibitemShut {NoStop}%
\bibitem [{\citenamefont {Kimura}\ \emph
  {et~al.}(2020{\natexlab{b}})\citenamefont {Kimura}, \citenamefont
  {Katsuyoshi}, \citenamefont {Sawada}, \citenamefont {Kimura},\ and\
  \citenamefont {Kimura}}]{Kimura20}%
  \BibitemOpen
  \bibfield  {author} {\bibinfo {author} {\bibfnamefont {K.}~\bibnamefont
  {Kimura}}, \bibinfo {author} {\bibfnamefont {T.}~\bibnamefont {Katsuyoshi}},
  \bibinfo {author} {\bibfnamefont {Y.}~\bibnamefont {Sawada}}, \bibinfo
  {author} {\bibfnamefont {S.}~\bibnamefont {Kimura}}, \ and\ \bibinfo {author}
  {\bibfnamefont {T.}~\bibnamefont {Kimura}},\ }\href {\doibase
  10.1038/s43246-020-0040-3} {\bibfield  {journal} {\bibinfo  {journal}
  {Commun. Mater.}\ }\textbf {\bibinfo {volume} {1}},\ \bibinfo {pages} {39}
  (\bibinfo {year} {2020}{\natexlab{b}})}\BibitemShut {NoStop}%
\bibitem [{\citenamefont {Akaki}\ \emph {et~al.}(2021)\citenamefont {Akaki},
  \citenamefont {Kimura}, \citenamefont {Kato}, \citenamefont {Sawada},
  \citenamefont {Narumi}, \citenamefont {Ohta}, \citenamefont {Kimura},
  \citenamefont {Motome},\ and\ \citenamefont {Hagiwara}}]{Akaki21}%
  \BibitemOpen
  \bibfield  {author} {\bibinfo {author} {\bibfnamefont {M.}~\bibnamefont
  {Akaki}}, \bibinfo {author} {\bibfnamefont {K.}~\bibnamefont {Kimura}},
  \bibinfo {author} {\bibfnamefont {Y.}~\bibnamefont {Kato}}, \bibinfo {author}
  {\bibfnamefont {Y.}~\bibnamefont {Sawada}}, \bibinfo {author} {\bibfnamefont
  {Y.}~\bibnamefont {Narumi}}, \bibinfo {author} {\bibfnamefont
  {H.}~\bibnamefont {Ohta}}, \bibinfo {author} {\bibfnamefont {T.}~\bibnamefont
  {Kimura}}, \bibinfo {author} {\bibfnamefont {Y.}~\bibnamefont {Motome}}, \
  and\ \bibinfo {author} {\bibfnamefont {M.}~\bibnamefont {Hagiwara}},\ }\href
  {\doibase 10.1103/PhysRevResearch.3.L042043} {\bibfield  {journal} {\bibinfo
  {journal} {Phys. Rev. Res.}\ }\textbf {\bibinfo {volume} {3}},\ \bibinfo
  {pages} {L042043} (\bibinfo {year} {2021})}\BibitemShut {NoStop}%
\bibitem [{\citenamefont {Kimura}\ \emph
  {et~al.}(2018{\natexlab{b}})\citenamefont {Kimura}, \citenamefont {Sera},
  \citenamefont {Nakano}, \citenamefont {Nozue},\ and\ \citenamefont
  {Kimura}}]{Kimura_2018_Physica_magneto}%
  \BibitemOpen
  \bibfield  {author} {\bibinfo {author} {\bibfnamefont {K.}~\bibnamefont
  {Kimura}}, \bibinfo {author} {\bibfnamefont {M.}~\bibnamefont {Sera}},
  \bibinfo {author} {\bibfnamefont {T.}~\bibnamefont {Nakano}}, \bibinfo
  {author} {\bibfnamefont {Y.}~\bibnamefont {Nozue}}, \ and\ \bibinfo {author}
  {\bibfnamefont {T.}~\bibnamefont {Kimura}},\ }\href {\doibase
  https://doi.org/10.1016/j.physb.2017.10.101} {\bibfield  {journal} {\bibinfo
  {journal} {Physica B}\ }\textbf {\bibinfo {volume} {536}},\ \bibinfo {pages}
  {93} (\bibinfo {year} {2018}{\natexlab{b}})}\BibitemShut {NoStop}%
\bibitem [{\citenamefont {Kimura}\ \emph {et~al.}(2019)\citenamefont {Kimura},
  \citenamefont {Kimura},\ and\ \citenamefont {Kimura}}]{Kimura_2019_JPS}%
  \BibitemOpen
  \bibfield  {author} {\bibinfo {author} {\bibfnamefont {K.}~\bibnamefont
  {Kimura}}, \bibinfo {author} {\bibfnamefont {S.}~\bibnamefont {Kimura}}, \
  and\ \bibinfo {author} {\bibfnamefont {T.}~\bibnamefont {Kimura}},\ }\href
  {\doibase 10.7566/JPSJ.88.093707} {\bibfield  {journal} {\bibinfo  {journal}
  {J. Phys. Soc. Jpn.}\ }\textbf {\bibinfo {volume} {88}},\ \bibinfo {pages}
  {093707} (\bibinfo {year} {2019})}\BibitemShut {NoStop}%
\bibitem [{\citenamefont {R\"asta}\ \emph {et~al.}(2020)\citenamefont
  {R\"asta}, \citenamefont {Heinmaa}, \citenamefont {Kimura}, \citenamefont
  {Kimura},\ and\ \citenamefont {Stern}}]{Rasta_2020}%
  \BibitemOpen
  \bibfield  {author} {\bibinfo {author} {\bibfnamefont {R.}~\bibnamefont
  {R\"asta}}, \bibinfo {author} {\bibfnamefont {I.}~\bibnamefont {Heinmaa}},
  \bibinfo {author} {\bibfnamefont {K.}~\bibnamefont {Kimura}}, \bibinfo
  {author} {\bibfnamefont {T.}~\bibnamefont {Kimura}}, \ and\ \bibinfo {author}
  {\bibfnamefont {R.}~\bibnamefont {Stern}},\ }\href {\doibase
  10.1103/PhysRevB.101.054417} {\bibfield  {journal} {\bibinfo  {journal}
  {Phys. Rev. B}\ }\textbf {\bibinfo {volume} {101}},\ \bibinfo {pages}
  {054417} (\bibinfo {year} {2020})}\BibitemShut {NoStop}%
\bibitem [{\citenamefont {Rodriguez}\ \emph {et~al.}(2008)\citenamefont
  {Rodriguez}, \citenamefont {Adler}, \citenamefont {Brand}, \citenamefont
  {Broholm}, \citenamefont {Cook}, \citenamefont {Brocker}, \citenamefont
  {Hammond}, \citenamefont {Huang}, \citenamefont {Hundertmark},\ and\
  \citenamefont {Lynn}}]{Rodriguez08}%
  \BibitemOpen
  \bibfield  {author} {\bibinfo {author} {\bibfnamefont {J.~A.}\ \bibnamefont
  {Rodriguez}}, \bibinfo {author} {\bibfnamefont {D.~M.}\ \bibnamefont
  {Adler}}, \bibinfo {author} {\bibfnamefont {P.~C.}\ \bibnamefont {Brand}},
  \bibinfo {author} {\bibfnamefont {C.}~\bibnamefont {Broholm}}, \bibinfo
  {author} {\bibfnamefont {J.~C.}\ \bibnamefont {Cook}}, \bibinfo {author}
  {\bibfnamefont {C.}~\bibnamefont {Brocker}}, \bibinfo {author} {\bibfnamefont
  {R.}~\bibnamefont {Hammond}}, \bibinfo {author} {\bibfnamefont
  {Z.}~\bibnamefont {Huang}}, \bibinfo {author} {\bibfnamefont
  {P.}~\bibnamefont {Hundertmark}}, \ and\ \bibinfo {author} {\bibfnamefont
  {J.~W.}\ \bibnamefont {Lynn}},\ }\href {\doibase
  10.1088/0957-0233/19/3/034023} {\bibfield  {journal} {\bibinfo  {journal}
  {Meas. Sci. Technol.}\ }\textbf {\bibinfo {volume} {19}},\ \bibinfo {pages}
  {034023} (\bibinfo {year} {2008})}\BibitemShut {NoStop}%
\bibitem [{\citenamefont {Babkevich}\ \emph
  {et~al.}(2016{\natexlab{b}})\citenamefont {Babkevich}, \citenamefont {Testa},
  \citenamefont {R\o{}nnow}, \citenamefont {Kimura},\ and\ \citenamefont
  {Ollivier}}]{DOI_IN5}%
  \BibitemOpen
  \bibfield  {author} {\bibinfo {author} {\bibfnamefont {P.}~\bibnamefont
  {Babkevich}}, \bibinfo {author} {\bibfnamefont {L.}~\bibnamefont {Testa}},
  \bibinfo {author} {\bibfnamefont {H.~M.}\ \bibnamefont {R\o{}nnow}}, \bibinfo
  {author} {\bibfnamefont {K.}~\bibnamefont {Kimura}}, \ and\ \bibinfo {author}
  {\bibfnamefont {J.}~\bibnamefont {Ollivier}},\ }\href {\doibase
  doi:10.5291/ILL-DATA.4-01-1528} {\  (\bibinfo {year} {2016}{\natexlab{b}}),\
  doi:10.5291/ILL-DATA.4-01-1528}\BibitemShut {NoStop}%
\bibitem [{\citenamefont {Testa}\ \emph {et~al.}(2019)\citenamefont {Testa},
  \citenamefont {R\o{}nnow},\ and\ \citenamefont {Raymond}}]{DOI_IN12}%
  \BibitemOpen
  \bibfield  {author} {\bibinfo {author} {\bibfnamefont {L.}~\bibnamefont
  {Testa}}, \bibinfo {author} {\bibfnamefont {H.~M.}\ \bibnamefont
  {R\o{}nnow}}, \ and\ \bibinfo {author} {\bibfnamefont {S.}~\bibnamefont
  {Raymond}},\ }\href {\doibase doi:10.5291/ILL-DATA.CRG-2652} {\  (\bibinfo
  {year} {2019}),\ doi:10.5291/ILL-DATA.CRG-2652}\BibitemShut {NoStop}%
\bibitem [{\citenamefont {Ewings}\ \emph {et~al.}(2016)\citenamefont {Ewings},
  \citenamefont {Buts}, \citenamefont {Le}, \citenamefont {{van Duijn}},
  \citenamefont {Bustinduy},\ and\ \citenamefont {Perring}}]{Ewings_2016}%
  \BibitemOpen
  \bibfield  {author} {\bibinfo {author} {\bibfnamefont {R.~A.}\ \bibnamefont
  {Ewings}}, \bibinfo {author} {\bibfnamefont {A.}~\bibnamefont {Buts}},
  \bibinfo {author} {\bibfnamefont {M.~D.}\ \bibnamefont {Le}}, \bibinfo
  {author} {\bibfnamefont {J.}~\bibnamefont {{van Duijn}}}, \bibinfo {author}
  {\bibfnamefont {I.}~\bibnamefont {Bustinduy}}, \ and\ \bibinfo {author}
  {\bibfnamefont {T.~G.}\ \bibnamefont {Perring}},\ }\href {\doibase
  https://doi.org/10.1016/j.nima.2016.07.036} {\bibfield  {journal} {\bibinfo
  {journal} {Nucl. Instrum. Methods Phys. Res., Sect. A}\ }\textbf {\bibinfo
  {volume} {834}},\ \bibinfo {pages} {132 } (\bibinfo {year}
  {2016})}\BibitemShut {NoStop}%
\bibitem [{\citenamefont {Toth}\ and\ \citenamefont {Lake}(2015)}]{spinw}%
  \BibitemOpen
  \bibfield  {author} {\bibinfo {author} {\bibfnamefont {S.}~\bibnamefont
  {Toth}}\ and\ \bibinfo {author} {\bibfnamefont {B.}~\bibnamefont {Lake}},\
  }\href {\doibase 10.1088/0953-8984/27/16/166002} {\bibfield  {journal}
  {\bibinfo  {journal} {J. Phys. Condens. Matter}\ }\textbf {\bibinfo {volume}
  {27}},\ \bibinfo {pages} {166002} (\bibinfo {year} {2015})}\BibitemShut
  {NoStop}%
\bibitem [{\citenamefont {Igarashi}(1992)}]{Igarashi92}%
  \BibitemOpen
  \bibfield  {author} {\bibinfo {author} {\bibfnamefont {J.-i.}\ \bibnamefont
  {Igarashi}},\ }\href {\doibase 10.1103/PhysRevB.42.10763} {\bibfield
  {journal} {\bibinfo  {journal} {Phys. Rev. B}\ }\textbf {\bibinfo {volume}
  {46}},\ \bibinfo {pages} {10763} (\bibinfo {year} {1992})}\BibitemShut
  {NoStop}%
\bibitem [{\citenamefont {Syromyatnikov}(2018)}]{Syromyatnikov2018}%
  \BibitemOpen
  \bibfield  {author} {\bibinfo {author} {\bibfnamefont {A.~V.}\ \bibnamefont
  {Syromyatnikov}},\ }\href {\doibase 10.1103/PhysRevB.98.184421} {\bibfield
  {journal} {\bibinfo  {journal} {Phys. Rev. B}\ }\textbf {\bibinfo {volume}
  {98}},\ \bibinfo {pages} {184421} (\bibinfo {year} {2018})}\BibitemShut
  {NoStop}%
\bibitem [{\citenamefont {Syromyatnikov}(2020)}]{Syromyatnikov2020}%
  \BibitemOpen
  \bibfield  {author} {\bibinfo {author} {\bibfnamefont {A.~V.}\ \bibnamefont
  {Syromyatnikov}},\ }\href {\doibase 10.1103/PhysRevB.102.014409} {\bibfield
  {journal} {\bibinfo  {journal} {Phys. Rev. B}\ }\textbf {\bibinfo {volume}
  {102}},\ \bibinfo {pages} {014409} (\bibinfo {year} {2020})}\BibitemShut
  {NoStop}%
\bibitem [{\citenamefont {Moretti~Sala}\ \emph {et~al.}(2015)\citenamefont
  {Moretti~Sala}, \citenamefont {Schnells}, \citenamefont {Boseggia},
  \citenamefont {Simonelli}, \citenamefont {Al-Zein}, \citenamefont {Vale},
  \citenamefont {Paolasini}, \citenamefont {Hunter}, \citenamefont {Perry},
  \citenamefont {Prabhakaran}, \citenamefont {Boothroyd}, \citenamefont
  {Krisch}, \citenamefont {Monaco}, \citenamefont {R\o{}nnow}, \citenamefont
  {McMorrow},\ and\ \citenamefont {Mila}}]{Moretti_Sala_2015}%
  \BibitemOpen
  \bibfield  {author} {\bibinfo {author} {\bibfnamefont {M.}~\bibnamefont
  {Moretti~Sala}}, \bibinfo {author} {\bibfnamefont {V.}~\bibnamefont
  {Schnells}}, \bibinfo {author} {\bibfnamefont {S.}~\bibnamefont {Boseggia}},
  \bibinfo {author} {\bibfnamefont {L.}~\bibnamefont {Simonelli}}, \bibinfo
  {author} {\bibfnamefont {A.}~\bibnamefont {Al-Zein}}, \bibinfo {author}
  {\bibfnamefont {J.~G.}\ \bibnamefont {Vale}}, \bibinfo {author}
  {\bibfnamefont {L.}~\bibnamefont {Paolasini}}, \bibinfo {author}
  {\bibfnamefont {E.~C.}\ \bibnamefont {Hunter}}, \bibinfo {author}
  {\bibfnamefont {R.~S.}\ \bibnamefont {Perry}}, \bibinfo {author}
  {\bibfnamefont {D.}~\bibnamefont {Prabhakaran}}, \bibinfo {author}
  {\bibfnamefont {A.~T.}\ \bibnamefont {Boothroyd}}, \bibinfo {author}
  {\bibfnamefont {M.}~\bibnamefont {Krisch}}, \bibinfo {author} {\bibfnamefont
  {G.}~\bibnamefont {Monaco}}, \bibinfo {author} {\bibfnamefont {H.~M.}\
  \bibnamefont {R\o{}nnow}}, \bibinfo {author} {\bibfnamefont {D.~F.}\
  \bibnamefont {McMorrow}}, \ and\ \bibinfo {author} {\bibfnamefont
  {F.}~\bibnamefont {Mila}},\ }\href {\doibase 10.1103/PhysRevB.92.024405}
  {\bibfield  {journal} {\bibinfo  {journal} {Phys. Rev. B}\ }\textbf {\bibinfo
  {volume} {92}},\ \bibinfo {pages} {024405} (\bibinfo {year}
  {2015})}\BibitemShut {NoStop}%
\bibitem [{\citenamefont {Katsura}\ \emph {et~al.}(2010)\citenamefont
  {Katsura}, \citenamefont {Nagaosa},\ and\ \citenamefont {Lee}}]{Katsura10}%
  \BibitemOpen
  \bibfield  {author} {\bibinfo {author} {\bibfnamefont {H.}~\bibnamefont
  {Katsura}}, \bibinfo {author} {\bibfnamefont {N.}~\bibnamefont {Nagaosa}}, \
  and\ \bibinfo {author} {\bibfnamefont {P.~A.}\ \bibnamefont {Lee}},\ }\href
  {\doibase 10.1103/PhysRevLett.104.066403} {\bibfield  {journal} {\bibinfo
  {journal} {Phys. Rev. Lett.}\ }\textbf {\bibinfo {volume} {104}},\ \bibinfo
  {pages} {066403} (\bibinfo {year} {2010})}\BibitemShut {NoStop}%
\bibitem [{\citenamefont {Shindou}\ \emph {et~al.}(2013)\citenamefont
  {Shindou}, \citenamefont {Matsumoto}, \citenamefont {Murakami},\ and\
  \citenamefont {Ohe}}]{Shindou13}%
  \BibitemOpen
  \bibfield  {author} {\bibinfo {author} {\bibfnamefont {R.}~\bibnamefont
  {Shindou}}, \bibinfo {author} {\bibfnamefont {R.}~\bibnamefont {Matsumoto}},
  \bibinfo {author} {\bibfnamefont {S.}~\bibnamefont {Murakami}}, \ and\
  \bibinfo {author} {\bibfnamefont {J.-i.}\ \bibnamefont {Ohe}},\ }\href
  {\doibase 10.1103/PhysRevB.87.174427} {\bibfield  {journal} {\bibinfo
  {journal} {Phys. Rev. B}\ }\textbf {\bibinfo {volume} {87}},\ \bibinfo
  {pages} {174427} (\bibinfo {year} {2013})}\BibitemShut {NoStop}%
\bibitem [{\citenamefont {Mook}\ \emph {et~al.}(2014)\citenamefont {Mook},
  \citenamefont {Henk},\ and\ \citenamefont {Mertig}}]{Mook14}%
  \BibitemOpen
  \bibfield  {author} {\bibinfo {author} {\bibfnamefont {A.}~\bibnamefont
  {Mook}}, \bibinfo {author} {\bibfnamefont {J.}~\bibnamefont {Henk}}, \ and\
  \bibinfo {author} {\bibfnamefont {I.}~\bibnamefont {Mertig}},\ }\href
  {\doibase 10.1103/PhysRevB.90.024412} {\bibfield  {journal} {\bibinfo
  {journal} {Phys. Rev. B}\ }\textbf {\bibinfo {volume} {90}},\ \bibinfo
  {pages} {024412} (\bibinfo {year} {2014})}\BibitemShut {NoStop}%
\bibitem [{\citenamefont {{van Hoogdalem}}\ \emph {et~al.}(2013)\citenamefont
  {{van Hoogdalem}}, \citenamefont {Tserkovnyak},\ and\ \citenamefont
  {Loss}}]{vanHoogdalem13}%
  \BibitemOpen
  \bibfield  {author} {\bibinfo {author} {\bibfnamefont {K.~A.}\ \bibnamefont
  {{van Hoogdalem}}}, \bibinfo {author} {\bibfnamefont {Y.}~\bibnamefont
  {Tserkovnyak}}, \ and\ \bibinfo {author} {\bibfnamefont {D.}~\bibnamefont
  {Loss}},\ }\href {\doibase 10.1103/PhysRevB.87.024402} {\bibfield  {journal}
  {\bibinfo  {journal} {Phys. Rev. B}\ }\textbf {\bibinfo {volume} {87}},\
  \bibinfo {pages} {024402} (\bibinfo {year} {2013})}\BibitemShut {NoStop}%
\bibitem [{\citenamefont {Romh{\'a}nyi}\ \emph {et~al.}(2015)\citenamefont
  {Romh{\'a}nyi}, \citenamefont {Penc},\ and\ \citenamefont
  {Ganesh}}]{Romhanyi15}%
  \BibitemOpen
  \bibfield  {author} {\bibinfo {author} {\bibfnamefont {J.}~\bibnamefont
  {Romh{\'a}nyi}}, \bibinfo {author} {\bibfnamefont {K.}~\bibnamefont {Penc}},
  \ and\ \bibinfo {author} {\bibfnamefont {R.}~\bibnamefont {Ganesh}},\ }\href
  {\doibase 10.1038/ncomms7805} {\bibfield  {journal} {\bibinfo  {journal}
  {Nat. Commun.}\ }\textbf {\bibinfo {volume} {6}},\ \bibinfo {pages} {6805}
  (\bibinfo {year} {2015})}\BibitemShut {NoStop}%
\end{thebibliography}%

\end{document}